\documentclass[12pt,12pt]{article}
\usepackage{amsmath,amssymb}
\usepackage{slashed} 

\catcode`\@=11
\newdimen\ex@
\def\beq{\begin{equation}}
\def\eeq{\end{equation}}
\def\beqa{\begin{eqnarray}}
\def\eeqa{\end{eqnarray}}
\newcommand{\ba}{\begin{eqnarray}}
\newcommand{\ea}{\end{eqnarray}}
\newcommand\BA{\begin{array}}
\newcommand\EA{\end{array}}

\begin{document}
\def\thefootnote{\fnsymbol{footnote}}

\title{\bf
$\bf \frac{SO(2N)}{U(N)}$ Riccati-Hartree-Bogoliubov Equation\\
Based on the $\bf SO(2N)$ Lie Algebra\\
of the Fermion Operators}
\vskip0.4cm
\author{Seiya NISHIYAMA$\!$\footnotemark[1]~~and
Jo\~ao da PROVID\^{E}NCIA$\!$\footnotemark[2]\\ 
\\
\\[-0.4cm]
Centro de F\'\i sica,
Departamento de F\'\i sica,\\
Universidade de Coimbra,
P-3004-516 Coimbra, Portugal
\\ \\[-0.4cm]
{\it Dedicated to the Memory of Hideo Fukutome}}

\maketitle

\vskip0.35cm

\footnotetext[1]
{Corresponding author.
~E-mail address: seikoceu@khe.biglobe.ne.jp,~
nisiyama@teor.fis.uc.pt}
\footnotetext[2]
{E-mail address: providencia@teor.fis.uc.pt}

\vspace{-1.5cm}

\begin{abstract}
$\!$In this paper
we present the induced representation of
$SO(2N)$ canonical transformation group
and introduce $\frac{SO(2N)}{U(N)}$ coset variables.
$\!\!$We give a derivation of the time dependent Hartree-Bogoliubov (TDHB) equation
on the K\"{a}hler coset space
$\!\frac{G}{H} \!\!=\!\! \frac{SO(2N)}{U(N)} \!$
from the Euler-Lagrange equation of motion for the coset variables.
The TDHB wave function represents
the TD behavior of Bose condensate of fermion pairs.
It is a good approximation for
the ground state of the fermion system with a pairing interaction,
producing the spontaneous Bose condensation.
To describe the classical motion on the coset manifold,
we start from the local equation of motion.
This equation becomes a Riccati-type equation.
After giving a simple two-level model and a solution for a coset variable,
we can get successfully a general solution of TDRHB equation
for the coset variables.
We obtain the Harish-Chandra decomposition
for the $SO(2N)$ matrix
based on the nonlinear M\"{o}bius transformation together with
the geodesic flow on the manifold.
\end{abstract}
\vspace{-0.2cm}
~~~~~~~Keywords: Kaehler manifold; Hartree-Bogoliubov theory; 
SO(2N) Lie algebra; 

~~~time dependent Riccati-Hartree-Bogoliubov equation \\
\vspace{0.1cm} 
 ~~~~~~Mathematics Subject Classification 2010: 81Rxx, 81R05, 81Vxx, 81V35


\newpage

\setcounter{equation}{0}
\renewcommand{\theequation}{\arabic{section}.\arabic{equation}}

\section{Introduction}
\vspace{-0.3cm}
~~
The supersymmetric (SUSY) extension of the nonlinear $\sigma$-model
was first given by Zumino under the introduction of scalar fields
on a K\"{a}hler manifold
\cite{Zumino.79}.
The extended $\sigma$-model defined on symmetric spaces
have been intensively studied
in modern elementary particle physics,
superstring theory and supergravity theory
\cite{NNH.01}.
While the Hartree-Bogoliubov theory (HBT)
\cite{Bog.59}
has been regarded as the standard approximation
in the theory of fermion systems
\cite{RS.80,BlaizotRipka.86}.
In HBT
a HB wave function (WF) represents
a Bose condensate of fermion pairs.
From the Lie-algebraic viewpoint,
fermion pair operators form a $SO(2N)$ Lie algebra
and contain a $U(N)$ Lie algebra as a subalgebra
($N$: Number of fermion states).
The $SO(2N)$ and $U(N)$ mean the special orthogonal group
of $2N$ dimensions and unitary group of $N$ dimensions, respectively.
One can give the exact coherent state representation (CS rep) of a fermion system
\cite{Perelomov.86}.

A consistent coupling of gauge- and matter superfields
to SUSY $\sigma$-models on K\"{a}hler coset space
has been given by van Holten et al.
$\!\!$They have presented a mathematical tool of constructing
a Killing potential and have applied it
to the explicit construction of a SUSY $\sigma$-model
on the coset space $\frac{SO(2N)}{U(N)}$.
Fukutome et. al. have proposed a new fermion many-body theory
based on a $SO(2N \!\!+\!\! 1)$ Lie algebra of fermion operators
\cite{FYN.77}.
A rep of the $SO(2N \!\!+\!\! 1)$ group has been derived
by a group extension of a $SO(2N)$ Bogoliubov transformation
for fermions to a new canonical transformation group.
The fermion Lie operators are represented by bosons
\cite{YN.76}.

We have extended the SUSY $\sigma$-model
on the K\"{a}hler coset space
$\!\frac{G}{H} \!=\!\! \frac{SO(2N \!+\! 2)}{U(N \!+\! 1)}\!$
based on the $SO(2N \!\!+\!\! 1)$ Lie algebra of fermion operators
\cite{SJCF.08}.
Following Fukutome,
by embedding a $\!SO(2N \!\!+\!\! 1)\!$ group into a $\!SO(2\!N \!+\! 2)\!$ group and
using $\!\frac{SO(2N \!+\! 2)}{U(N \!+\! 1)}\!$ coset variables
\cite{Fuk.77,Fuk.81},
we have studied a new aspect of the extended SUSY $\sigma$-model
and then constructed a Killing potential.
Using the coset space $\frac{SO(2N)}{U(N)}$
and constructing the K\"{a}hler and Killing potentials, 
van Holten et. al. have provided
the anomaly-free SUSY $\sigma$-model on $\frac{SO(2N)}{U(N)}$
\cite{NNH.01}.
The time dependent (TD) HBT is
a powerful tool for describing superconducting fermion systems
\cite{Bog.59,RS.80}. 
The TDHB WF represents
the TD behavior of Bose condensate states of fermion pairs.
It is a good approximation for
the ground state of a fermion system with a pairing interaction,
producing the spontaneous Bose condensation.
The TDHB equation on the K\"{a}hler coset space
$\!\frac{G}{H} \!=\!\! \frac{SO(2N)}{U(N)} \!$ is derived
from the Euler-Lagrange
equation of motion for the coset variables.
To describe the classical motion on the coset manifold,
we start from the local equation of motion
\cite{Nishi.81,BG.91}.
This equation becomes of the 
\cite{Riccati.1758,Reid.72,Z_Itkin.73}.
After providing a simple two-level model and a solution for a coset variable,
we give a general solution of the TDRHB equation
for the coset variables.
We obtain the Harish-Chandra decomposition
for the $SO(2N)$ matrix
based on the nonlinear M\"{o}bius transformation together with
the geodesic flow on the manifold.

This paper is organized as follows:
In \S 2,
we recapitulate briefly the
induced representation of the $SO(2N)$ canonical transformation group
and the introduction of the $\frac{SO(2N)}{U(N)}$ coset variables.
We give a brief sketch of
the derivation of the TDHB and the TDRHB equations from the classical
Euler-Lagrange equation of motion for the $\frac{SO(2N)}{U(N)}$
coset variables in the TD self-consistent field (TDSCF).
In \S 3,
we give the $\frac{SO(2N)}{U(N)}$
K\"{a}hler and Killing potentials
and present the Harish-Chandra decomposition
for the $SO(2N)$ matrix
based on the nonlinear M\"{o}bius transformation
and the Killing potential for tensors.
In \S 4, we provide a simple two-level model and
a solution for a coset variable
up to the first order and the infinite order in time $t$.
In \S 5, 
we give a general solution of the TDRHB equation
for the $\frac{SO(2N)}{U(N)}$ coset variables.
Finally, in the last section, we give some concluding remarks
and further outlook.
In the Appendices,
after providing the bosonization of $SO(2N)$ Lie operators and
vacuum function for bosons,
we give another way of derivation of the TDRHB equation. 
Finally, 
we give the TDRHB equations for three- and four- level models.


\newpage

\setcounter{equation}{0}
\renewcommand{\theequation}{\arabic{section}.\arabic{equation}}

\section{Brief summary of $\bf SO(2N)$ Bogoliubov transformation and Derivation of
$\bf \frac{SO(2N)}{U(N)}$ Riccati-Hartree-Bogoliubov equation}

\def\bra#1{{<\!#1\,|}} 
\def\ket#1{{|\,#1\!>}}

~
We give a brief summary
of a $SO(2N)$ canonical transformation
and of fixing a $\!\frac{SO(2N)}{U(N)}$ coset variable.
\def\bra#1{{<\!#1\,|}} 
\def\ket#1{{|\,#1\!>}}
$\!\!$Let $c_{\alpha }$ and $c^{\dag }_{\alpha }$,
$\!(
\alpha
\!=\! 
1, \!\cdots\!, N
)$, 
be annihilation and creation operators of the fermion
satisfying the canonical anti-commutation relations~
$
\{c_{\alpha }, c^{\dag }_{\beta }\}
\!=\! 
\delta_{\alpha \beta } ,
\{c_{\alpha }, c_{\beta }\}
\!=\!
0
$
and
$ 
\{c^{\dag }_{\alpha }, c^{\dag }_{\beta }\}
\!=\! 
0 
$.
We introduce the set of fermion operators consisting of the following
pair operators:\\[-22pt] 
\beqa
\left.
\BA{ll}
&E^{\alpha }_{~\beta }
=
c^{\dag }_{\alpha }c_{\beta }
{\displaystyle -\frac{1}{2}} \delta_{\alpha \beta } ,~~
E^{\alpha \beta }
=
c^{\dag }_{\alpha }c^{\dag }_{\beta } ,~~
E_{\alpha \beta }
=
c_{\alpha }c_{\beta } ,\\
\\[-10pt]
&E^{\alpha \dag }_{~\beta }
=
E^{\beta }_{~\alpha } ,~~
E^{\alpha \beta } 
=
E^{\dag }_{\beta \alpha } ,~~
E_{\alpha \beta }
=
- E_{\beta \alpha } .~~
(\alpha , \beta = 1, \cdot \cdot \cdot, N)
\EA
\right\}
\label{operatorset}
\eeqa\\[-14pt]
It is well known that the set of fermion operators
(\ref{operatorset})
form a $SO(2N)$ Lie algebra.
As a consequence of the anti-commutation relations,
the commutation relations for the fermion operators
(\ref{operatorset})
in the $SO(2N)$ Lie algebra are\\[-22pt] 
\beqa
[E^{\alpha }_{~\beta },~E^{\gamma }_{~\delta }]
=
\delta_{\gamma \beta }E^{\alpha }_{~\delta } 
- 
\delta_{\alpha \delta }E^{\gamma }_{~\beta },~~
(U(N)~\mbox{algebra})
\label{commurel1}
\eeqa
\vspace{-1.0cm}
\beqa
\left.
\BA{ll}
&[E^{\alpha }_{~\beta },~E_{\gamma \delta }]
=
\delta_{\alpha \delta }E_{\beta \gamma } 
- 
\delta_{\alpha \gamma }E_{\beta \delta },\\
\\[-6pt]
&[E^{\alpha \beta },~E_{\gamma \delta }]
=
\delta_{\alpha \delta }E^{\beta }_{~\gamma } 
+ 
\delta_{\beta \gamma }E^{\alpha }_{~\delta }
-
\delta_{\alpha \gamma }E^{\beta }_{~\delta } 
- 
\delta_{\beta \delta }E^{\alpha }_{~\gamma },~~
[E_{\alpha \beta },~E_{\gamma \delta }]
=
0,
\EA
\right\}
\label{commurel2}
\eeqa\\[-14pt]
We omit the commutation relations obtained 
by hermitian conjugation of
(\ref{commurel2}).

A $SO(2N)$ canonical transformation $U(g)$ is generated by 
the fermion $SO(2N)$ Lie operators.
The transformation $U(g)$
is expressed by a successive transformation as
$U(g) \!=\! e^{\Lambda}e^{\Gamma} $,
$
\Gamma
\!\!=\!\! 
\gamma_{\alpha \beta }
c^{\dag }_{\alpha }c_{\beta }
(\bar{\gamma}^{\dag } \!\!=\!\! - \gamma)
$
and
$
\Lambda
\!\!=\!\!
{\displaystyle \frac{1}{2}} \!
\left( \!
\lambda_{\alpha \beta }c^{\dag }_{\alpha }c^{\dag }_{\beta }
\!+\!
\bar{\lambda}_{\alpha \beta }c_{\alpha }c_{\beta } \!
\right) \! 
(\lambda^{\mbox{\scriptsize T}} \!\!=\!\! - \lambda)
$.
Introduce matrices $g_\gamma$ and $g_\lambda$ as
$
g_\gamma
\!\!=\!\!
\left[ \!\!\!
\BA{cc}
\bar{\gamma} & 0 \\
\\[-10pt]
0 &\!\!\!\! \gamma \!\!\!\!\!
\EA \!\!
\right]
$
and
$
g_\lambda
\!\!=\!\!
\left[ \!\!\!
\BA{cc}
C(\lambda) &\!\!\!\! \bar{S}(\lambda) \\
\\[-10pt]
S(\lambda) &\!\!\!\!\!\!\!\! \bar{C}(\lambda) \!\!\!\!\!\!
\EA \!\!
\right] 
$. 
Then the $U(g)$ is the generalized Bogoliubov transformation 
\cite{Bog.59} 
specified by an $SO(2N)$ matrix $g~(\det g \!=\! 1)$ as, \\[-10pt]
\beq
U(g)(c, c^{\dag })U^{\dag }(g)
\!=\!
(c, c^{\dag }) g ,~
g
\!=\!
g_\lambda g_\gamma
\!\equiv\!\! 
\left[ \!\!\!
\BA{cc}
a & \bar{b} \\
\\[-10pt]
b &\!\! \bar{a} \!\!\!
\EA \!\!
\right] \! ,
\BA{c}
\!\!
a
\!=\!
C(\lambda) \bar{\gamma},
C(\lambda)
\!\equiv\!
\cos( \sqrt{\lambda^\dag \lambda} ) , \\
\\[-16pt]
~
b
\!=\!
S(\lambda) \bar{\gamma},~
S(\lambda)
\!\equiv\!
\lambda 
{\displaystyle \frac{\sin( \sqrt{\lambda^\dag \lambda} )}{\sqrt{\lambda^\dag \lambda}}},
\EA
\label{Bogotrans}
\eeq\\[-14pt]
and
$g^{\dag }g \!=\! gg^{\dag } \!=\! 1_{\!2N}$.
The $U(g)$ also satisfies the following properties:\\[-10pt]
\beq
U(g)U(g') \!=\! U(gg') ,~~
U(g^{-1}) \!=\! U^{-1}(g) \!=\! U^{\dag }(g) ,~~
U(1_{\!2N}) \!=\! \mathbb{I} .
\label{Ug}
\eeq
Here, ($c$, $c^{\dag }$) is
a 2$N$-dimensional row vector
(($c_{\alpha }$), ($c^{\dag }_{\alpha }$)) and both the
$a \!=\! (a^{\alpha }_{~\beta })$ and $b \!=\! (b_{\alpha \beta })$
are $N \!\times\! N$ matrices.
The HB ($SO(2N)$) wave function $\ket g$ is defined as
$\ket g \!=\! U(g) \ket 0$ 
($\ket 0$ : the vacuum satisfying 
$c_{\alpha }\ket 0 \!=\! 0$).
The wave function, coherent state $\ket g$,  is expressed as\\[-22pt]
\beqa
\ket g
\!=\!
\bra 0 U(g) \ket 0
\exp({\displaystyle \frac{1}{2}} q_{\alpha \beta }c^\dagger_\alpha 
c^\dagger_\beta) \ket 0 ,
\label{Bogoketg}
\eeqa
\vspace{-1.0cm}
\beqa
\bra 0 U(g) \ket 0
\!=\!
\left[\det(a)\right]^{\frac{1}{2}}
\!=\!
\left[\det(1_{\!N} \!+\! q^\dag q)\right]^{-\frac{1}{4}}
e^{i \frac{\tau}{2} } ,~
\tau
\!=\!
{\displaystyle \frac{i}{2}} \ln \!
\left[{\displaystyle \frac{\det({\bar{a}})}{\det({a})}} \!
\right]  ,
\label{Bogowf}
\eeqa
\vspace{-1.0cm}
\beqa
q \!=\! ba^{-1}
\!=\!
\lambda 
{\displaystyle \frac{\tan( \sqrt{\lambda^\dag \lambda})}{\sqrt{\lambda^\dag \lambda}}} ,~
q \!=\! -q^{\mbox{\scriptsize T}} .
\label{Bogocoset}
\eeqa\\[-12pt]
The $q$ is a variable of the $\frac{SO(2N)}{U(N)}$ coset space.
The $\tau$ is a phase of the $U(N)$ subgroup and
the derivative ${\displaystyle \frac{\partial }{\partial \tau }}$ appeared in
(\ref{differentialformulas})
in Appendix A 
plays a crucial role to show the existence of the free fermion vacuum.
The function $\overline{\bra \! 0 U(g)  \! \ket 0}~(\equiv\! \Phi_{00}(g))$
in $g \!\in\! SO(2N)$ corresponds to the free fermion vacuum function
as proved in Appendix B.
The symbols $\det$ and {\scriptsize T} denote
the determinant and transposition, respectively.
The overline denotes the complex conjugation.

\def\erw#1{{<\!\!#1\!\!>_g}}
Expectation values of the fermion
$SO (2N)$ Lie operators, i.e.,
the generators of rotation
in $2N$-dimensional Euclidian space,
with respect to $| g \!\! >$ are given as\\[-14pt]
\ba
\left.
\BA{ll}
&\erw{E^\alpha_{~\beta }
\!+\!
{\displaystyle \frac{1}{2}\delta_{\alpha \beta }}}
\!=\!
R_{\alpha \beta }
\!=\!
{\displaystyle \frac{1}{2}} \!
\left( \!
\bar{b}_{\alpha i} b_{\beta i}
\!-\!
a^\alpha _{~i} \bar{a}^{\beta }_{~i}
\right)
\!+\!
{\displaystyle \frac{1}{2}}\delta_{\alpha \beta }
\!=\!
-
\left[ \bar{q} q (1_{\!N} \!-\!  \bar{q} q)^{-1} \right]_{\alpha \beta } ,\\
\\[-16pt]
&\erw{E_{\alpha \beta }}
\!=\!
-K_{\alpha \beta }
\!=\!
{\displaystyle \frac{1}{2}} \!
\left( \!
\bar{a}^{\alpha }_{~i} b_{\beta i}
\!-\!
b_{\alpha i} \bar{a}^{ \beta }_{~i}
\right)
\!=\!
-
\left[ q (1_{\!N} \!-\!  \bar{q} q)^{-1} \right]_{\alpha \beta } ,~
\erw{E^{\alpha \beta }}
\!=\!
\bar{K}_{\alpha \beta } .
\EA
\right\}
\label{expectG}
\ea
The expectation value of a two-body operator is given as
\\[-14pt]
\ba
\erw{E^{\alpha\gamma } \! E_{\delta\beta }}
\!=\!
R_{\alpha\beta }R_{\gamma\delta }-
R_{\alpha\delta }R_{\gamma\beta }-
\bar{K}_{\alpha\gamma }K_{\delta\beta } .
\label{2bodtexpectG}
\ea

Let the Hamiltonian of the fermion system
under consideration be\\[-14pt]
\ba
H
\!=\!
h_{\alpha\beta } \!
\left( \! E^\alpha_{~\beta } \!+\! \frac{1}{2}\delta_{\alpha\beta } \! \right)
\!+\!
\frac{1}{4}[\alpha\beta|\gamma\delta]
E^{\alpha\gamma }E_{\delta\beta } .
\label{Hamiltonian}
\ea
The matrix $h_{\alpha\beta }$ related to a single-particle hamiltonian includes a chemical
potential and
$
[\alpha\beta|\gamma\delta]
\!=\!
-
[\alpha\delta|\gamma\beta]
\!=\!
[\gamma\delta|\alpha\beta]
\!=\!
\overline{[\beta\alpha|\delta\gamma]}
$
are anti-symmetrized matrix
elements of an interaction potential.
Parallel to calculations by the usual HB factorization method
(see Refs.\cite{RS.80}
and
\cite{BlaizotRipka.86}),
the expectation value of $H$ with respect to $| g \!\! >$
is calculated as\\[-14pt]
\ba
\BA{ll}
\erw H
&\!\!\!=
h_{\alpha\beta }\erw{E^\alpha_{~\beta }
\!\!+\!\!
{\displaystyle \frac{1}{2}}\delta_{\alpha\beta }}
\!\!+\!\!
{\displaystyle \frac{1}{2}}[\alpha\beta|\gamma\delta] \!
\left\{ \!\!
\erw{E^\alpha_{~\beta }
\!+\!
{\displaystyle \frac{1}{2}\delta_{\alpha\beta }}}
\erw{E^\gamma_{~\delta }
\!\!+\!\!
{\displaystyle \frac{1}{2}}\delta_{\gamma\delta }}
\!\!+\!\!
{\displaystyle \frac{1}{2}}\erw{E^{\alpha\gamma }}
\erw{E_{\delta\beta }} \!\!
\right\} \\
\\[-16pt]
&\!\!\!=
h_{\alpha\beta}R_{\alpha\beta}
+
{\displaystyle \frac{1}{2}}[\alpha\beta|\gamma\delta] \!
\left( \!
R_{\alpha\beta}R_{\gamma\delta}
-
{\displaystyle \frac{1}{2}}K^\star_{\alpha\gamma}K_{\delta\beta} \!
\right) .
\EA
\label{HexpectG}
\ea
The $SO(2N)$ TDHB equation can be derived from the Euler-Lagrange
equation of motion for the $\!\frac{SO(2N)}{U(N)}\!$ coset variables $\!q\!$
(\ref{Bogocoset}). 
We start from the local equations of motion
\cite{Nishi.81,BG.91}
given by\\[-14pt]
\ba
\left.
\BA{cc}
&\dot q
=
-{\displaystyle \frac{i}{\hbar}}(1_{\!N} \!-\! \bar{R})^{-1}
{\displaystyle \frac{\partial \erw H }{\partial  \bar{q}}}(1_{\!N} \!-\! R)^{-1} ,~
q (1_{\!N} \!-\! R) = K ,~ \bar{q} K = - R , \\
\\[-16pt]
&\dot{ \bar{q}}
=
-{\displaystyle \frac{i}{\hbar}}(1_{\!N} \!-\! R)^{-1}
{\displaystyle \frac{\partial \erw H }{\partial q}}(1_{\!N} \!-\! \bar{R})^{-1} ,~
(1_{\!N} \!-\! \bar{R}) q = K ,~\bar{K} q = - R .
\EA
\right\}
\label{Classicaleqofmotion}
\ea
With the use of the differential formulae,\\[-14pt]
\ba
\!\!\!\!\!\!\!\!
\left.
\BA{ll}
&{\displaystyle \frac{\partial R_{\gamma\delta} }{\partial q_{\alpha\beta}}}
\!=\!
-
\bar{K} _{\gamma\alpha}(1_{\!N} \!-\! R)_{\beta\delta}
\!+\!
\bar{K} _{\gamma\beta}(1_{\!N} \!-\! R)_{\alpha\delta} ,~
{\displaystyle \frac{\partial R_{\gamma\delta} }{\partial \bar{q} _{\alpha\beta}}}
\!=\!
-
(1_{\!N} \!-\! R) _{\gamma\alpha} K_{\beta\delta}
\!+\!
(1_{\!N} \!-\! R) _{\gamma\beta} K_{\alpha\delta} , \\
\\[-10pt]
&{\displaystyle \frac{\partial K_{\gamma\delta} }{\partial q_{\alpha\beta}}}
\!=\!
(1_{\!N} \!-\! \bar{R}) _{\gamma\alpha}(1_{\!N} \!-\! R)_{\beta\delta}
\!-\!
(1_{\!N} \!-\! \bar{R}) _{\gamma\beta}(1_{\!N} \!-\! R)_{\alpha\delta} ,~
{\displaystyle \frac{\partial K_{\gamma\delta} }{\partial \bar{q} _{\alpha\beta}}}
\!=\!
K_{\gamma\alpha} K_{\beta\delta}
\!-\!
K_{\gamma\beta} K_{\alpha\delta} .
\EA \!\!
\right\}
\label{Differentialformulae}
\ea\\[-6pt]
Let us define matrices $Q$, ${\cal F}_{\!F}$ and ${\cal F}_{\!D}$ as
$
Q
\!\equiv\!
\left[ \!\!
\BA{cc}
0_{N} & q \\
\\[-14pt]
\bar{q} & 0_{N} 
\EA \!\!
\right]
$,
$
{\cal F}_F
\!\equiv\!
\left[ \!\!
\BA{cc}
F & 0_{N} \\
\\[-14pt]
0_{N} & -\bar{F} 
\EA \!\!
\right]
$
and
$
{\cal F}_D
\!\equiv\!
\left[ \!\!
\BA{cc}
0_{N} & D \\
\\[-14pt]
-\bar{D} & 0_{N} 
\EA \!\!
\right]
$.
Then the classical equation of motion for the $\frac{SO(2N)}{U(N)}$
coset variables is calculated to be\\[-12pt]
\ba
\BA{ll}
i\hbar \!
\left[ \!\!
\BA{cc}
0_{N} & \dot q \\
\\[-8pt]
\dot{\bar{q}} & 0_{N} 
\EA \!\!
\right]
\!\!=\!
{\cal F}_{\!D}
\!+\!
{\cal F}_{\!F}
Q
\!-\!
Q
{\cal F}_{\!F}
\!-\!
Q
\cal{F}_{\!D}
Q
\!=\!\!
\left[ \!\!
\BA{cc}
0_{N} &\!\! D \!+\! F q \!+\! q \bar{F}\!+\! q \bar{D} q \\
\\[-8pt]
-\bar{D} \!-\! \bar{F} \bar{q} \!-\! \bar{q} F \!-\! \bar{q} D \bar{q} &\!\! 0_{N} 
\EA \!\!
\right] \! .
\EA
\label{RiccatiEq}
\ea\\[-8pt]
Equations
(\ref{RiccatiEq})
are just the Riccati-type matrix equations.
Then we call them
the  $\frac{SO(2N)}{U(N)}$ time dependent Riccati-Hartree-Bogoliubov (TDRHB) equations
and
${\cal F} (\!=\! {\cal F}_F \!+\! {\cal F}_D)$ is the Hartree-Bogoliubov
matrix Hamiltonian.
The SCF parameters $F \!=\! (F_{\alpha\beta}) \!=\! F^\dagger $ and
$D \!=\! (D_{\alpha\beta}) \!=\! -D^{\mbox{\scriptsize T}}$
appeared in the matrix elements in
(\ref{RiccatiEq})
are defined by the functional derivatives as\\[-14pt]
\ba
\BA{c}
F_{\alpha\beta}
\equiv
{\displaystyle \frac{\partial\erw H}{\partial R_{\alpha\beta}}}
=
h_{\alpha\beta}
+
[\alpha\beta|\gamma\delta] R_{\gamma\delta},~~
D_{\alpha\beta}
\equiv
{\displaystyle \frac{\partial\erw H }{\partial \bar{K}_{\alpha\beta}}}
=
{\displaystyle \frac{1}{2}}[\alpha\gamma|\beta\delta]
(-K_{\delta\gamma}) .
\EA
\label{SCFparameters}
\ea\\[-10pt] 
The classical equations of motion on the K\"{a}hlerian manifold
is put into the form
(\ref{Differentialformulae}).
The K\"{a}hlerian structure of the symmetric space is carried
onto the manifold of coherent state
\cite{BG.91}.

In the next Section, we study the K\"{a}hlerian structure through
the K\"{a}hler potential and the Killing potential.


\newpage

\setcounter{equation}{0}
\renewcommand{\theequation}{\arabic{section}.\arabic{equation}}

\section{$\bf \frac{SO(2N)}{U(N)}$ K\"{a}hler potential and Killing potential}
\vspace{-0.3cm} 
~~~
Let us introduce a $2N \!\times\! N$ isometric matrix 
$u$ by
$
u^{\mbox{\scriptsize T}}
=
\left[ 
\BA{cc} 
\!\!b^{\mbox{\scriptsize T}}, ~ a^{\mbox{\scriptsize T}}\!\!
\EA 
\right] 
$.
If one uses the matrix elements of 
$u$ and $u^\dag $
as the coordinates on the manifold $SO(2N)$,
a real line element can be defined by a hermitian metric tensor 
on the manifold.
Under the transformation
$u \!\rightarrow\! vu$
the metric is invariant.
Then the metric tensor defined on the manifold may become singular,
due to the fact that one uses too many coordinates.
 
According to Zumino
\cite{Zumino.79},
if $a$ is non-singular,
we have relations governing $u^\dag u$ as\\[-20pt]
\beqa
\left.
\BA{ll}
&
u^\dag u
\!=\!
a^\dag a
\!+\!
b^\dag b
\!=\!
a^\dag \!
\left\{
1_{\!N} 
\!+\!
\left(
ba^{-1} \right)^\dag \!
\left( ba^{-1} \right)
\right\} \!
a
\!=\!
a^\dag \!
\left(
1_{\!N} 
\!+\! 
q^\dag q
\right) \!
a ,\\
\\[-12pt]
&
\ln \det u^\dag u
\!=\!
\ln \det 
\left(
1_{\!N} 
\!+\! 
q^\dag q
\right)
\!+\!
\ln \det a
\!+\!
\ln \det a^\dag ,
\EA
\right\}
\label{lndetUUdagger}
\eeqa\\[-12pt]
where we have used 
the $\frac{SO(2N)}{U(N)}$ coset variable $q$
(\ref{Bogocoset}).
If we take the matrix elements of $q$ and $\bar{q}$ 
as the coordinates 
on the $\frac{SO(2N)}{U(N)}$ coset manifold,
the real line element can be well defined by a hermitian metric tensor 
on the coset manifold as\\[-14pt]
\beq
ds^2
\!=\!
G_{\alpha\beta \underline{\gamma}\underline{\delta}}
dq^{\alpha\beta}d\bar{q}^{\underline{\gamma}\underline{\delta}}~
(q^{\alpha\beta} = {q}_{\alpha\beta}~\mbox{and}~
\bar{q}^{\underline{\gamma}\underline{\delta}} 
\!=\!
\bar{q}_{\underline{\gamma}\underline{\delta}};~
G_{\alpha\beta \underline{\gamma}\underline{\delta}}
\!=\!
G_{\underline{\gamma}\underline{\delta} \alpha\beta}) .
\label{metric}
\eeq\\[-20pt]
We also use the indices
$\underline{\gamma},~\underline{\delta},~\cdots$ 
running over $\alpha,~\beta,~\cdots$.
The condition that the manifold under consideration is 
a K\"{a}hler manifold 
is that its complex structure is 
covariantly constant relative to the Riemann connection:\\[-12pt]
\beq
G_{\alpha\beta \underline{\gamma}\underline{\delta} , \epsilon\varphi}
\stackrel{\mathrm{def}}{=}
\frac{\partial G_{\alpha\beta \underline{\gamma}\underline{\delta}}}
{\partial q^{\epsilon\varphi}}
\!=\!
G_{\epsilon\varphi \underline{\gamma}\underline{\delta} ,\alpha\beta} ,~~
G_{\alpha\beta \underline{\gamma}\underline{\delta} , \underline{\epsilon} \underline{\varphi}}
\stackrel{\mathrm{def}}{=}
\frac{\partial G_{\alpha\beta \underline{\gamma}\underline{\delta}}}
{\partial \bar{q}^{\underline{\epsilon} \underline{\varphi}}}
\!=\!
G_{\alpha\beta \underline{\epsilon} \underline{\varphi} ,\underline{\gamma}\underline{\delta}} ,
\label{Kcondition}
\eeq\\[-16pt]
and that it has vanishing torsions.
Then, the hermitian metric tensor 
$G_{\alpha\beta \underline{\gamma}\underline{\delta}}$
can be locally given through a real scalar function,
the K\"{a}hler potential, 
which takes the well-known form\\[-14pt] 
\beq
{\cal K}(q^\dag, q) 
\!=\!
\ln \det 
\left(
1_{\!N}
\!+\! 
q^\dag q
\right) ,
\label{Kaehlerpotential}
\eeq\\[-20pt]
and the explicit expression for the components of the metric tensor
is given as\\[-24pt] 
\beqa
\!\!\!\!\!\!\!\!
\BA{ll}
G_{\alpha\beta \underline{\gamma}\underline{\delta}}
\!=\!
{\displaystyle 
\frac{\partial ^2 {\cal K}(q^\dag, q)}
{\partial q^{\alpha\beta} 
 \partial \bar{q}^{\underline{\gamma}\underline{\delta}}}
}
&\!\!\!
\!=\!
\left\{ \!\!
\left( \!
1_{\!N} 
\!+\! 
qq^\dag 
\right)^{-1} \!
\right\}_{\delta\alpha} \!\!
\left\{ \!\!
\left( \!
1_{\!N} 
\!+\! 
q^\dag q
\right)^{-1} \!
\right\}_{\beta\gamma} \!\!
\!-\!
(\gamma \!\leftrightarrow\! \delta)
\!-\! 
(\alpha \!\leftrightarrow\! \beta) 
\!+\!
(\alpha \!\leftrightarrow\! \beta, \gamma \!\leftrightarrow\! \delta) .
\EA
\label{metricfromKpot}
\eeqa\\[-18pt]
Notice that the above function does not determine
the K\"{a}hler potential 
${\cal K}(q^\dag, q)$ 
uniquely
since the metric tensor 
$G_{\alpha\beta \underline{\gamma}\underline{\delta}}$
is invariant under a transformation of the K\"{a}hler potential,\\[-14pt]
\beq
{\cal K}(q^\dag, q)
\!\rightarrow\!
{\cal K}^\prime (q^\dag, q)
\!=\!
{\cal K}(q^\dag, q)
\!+\! {\cal F}(q)
\!+\! \bar{\cal F}(\bar{q}) .
\label{transKpot}
\eeq\\[-20pt]
${\cal F}(q)$ and $\bar{\cal F}(\bar{q})$
are analytic functions of $q$ and $\bar{q}$, respectively.

Let us consider a $SO(2N)$ infinitesimal left transformation
of a $SO(2N)$ matrix $g$ to $g^\prime$,
$
g^\prime
=
(1_{\!2N} + \delta g) g
$,
by using the first equation of
(\ref{infinitesimalop}):\\[-24pt]
\beqa
g^\prime
\!=\!\! 
\left[ \!\!
\BA{cc}
1_{\!N} \!+\! \delta a & \delta \bar{b} \\
\\[-8pt]
\delta b & 1_{\!N} \!+\! \delta \bar{a} \!\!
\EA 
\right] \!\!
g
=\!
\left[ \!\!
\BA{cc}
a \!+\! \delta a a \!+\! \delta \bar{b} b  & 
\bar{b} \!+\! \delta a \bar{b} \!+\! \delta \bar{b} \bar{a} \\
\\[-8pt]
b \!+\! \delta \bar{a}{ b} \!+\! \delta b a  & 
\bar{a} \!+\! \delta \bar{a} \bar{a} \!+\! \delta b \bar{b} \!\!
\EA 
\right] .
\label{calGprime}
\eeqa\\[-16pt]
If $\delta a$ and $\delta b$ satisfy the relations
$
\delta a^\dag = - \delta a,
\mbox{tr}\delta a = 0~\mbox{and}~
\delta b = - \delta b^{\mbox{\scriptsize T}}
$,
the $(1_{\!2N} + \delta g)$
plays an important role to bosonize the $SO(2N)$ Lie operators
as presented in Appendix A.
Let us define a $\frac{SO(2N)}{U(N)}$ coset variable
$
q^\prime
(\!=\! b^\prime a^{\prime -1})
$
in the $g^\prime$ frame.
With the aid of 
(\ref{calGprime}),
the $q^\prime$ is calculated infinitesimally as\\[-26pt]
\beqa
\BA{ll}
q^\prime
=
b^\prime a^{\prime -1}
&\!\!\!
\!=\!
\left(
b \!+\! \delta \bar{a} b \!+\! \delta b a
\right) \!
\left(
a \!+\! \delta a a \!+\! \delta \bar{b} b
\right)^{-1} \\
\\[-10pt]
&\!\!\!
\!=\!
q \!+\! \delta b
\!-\! q \delta a \!+\! \delta \bar{a} q
\!-\! q \delta \bar{b} q .
\EA
\label{calQprime}
\eeqa\\[-30pt]

The K\"{a}hler metrics admit holomorphic isometries
(Killing vectors),
${\cal R}^{i\alpha}(q)$
and
$\bar{\cal R}^{i\underline{\alpha }}(\bar{q})
(i \!\!=\!\! 1, \cdots, \dim g, \alpha \!\!=\!\! 1,\cdots,N)$.
These isometries are solutions of the Killing equation\\[-16pt]
\beq
{\cal R}^i _{~\underline{\beta }}(q)_{,~\alpha}
\!+\!
\bar{\cal R}^i _{~\alpha}(q)_{,~\underline{\beta }}
=
0 ,~~
{\cal R}^i _{~\underline{\beta }}(q)
\!=\!
g_{\alpha \underline{\beta }}{\cal R}^{i\alpha}(q) .
\label{Killingeq}
\eeq\\[-18pt]
They define infinitesimal symmetry transformations and 
are described geometrically by the Killing vectors which
are generators of infinitesimal coordinate transformations
keeping the metric invariant: 
$
\delta q
\!=\!
q^\prime \!-\! q
\!=\!
{\cal R}(q)
$
and
$
\delta \bar{q}
\!=\!
\bar{\cal R}(\bar{q})
$
such that
$
g^\prime (q, \bar{q})
\!=\!
g (q, \bar{q})
$.
The Killing equation
is the necessary and sufficient condition for 
an infinitesimal coordinate transformation\\[-12pt]
\beq
\delta{q}^{\alpha}
\!=\!
\left(
\delta b 
\!-\! \delta a^{\mbox{\scriptsize T}}q \!-\! q \delta a
\!+\! q \delta b^\dag q
\right)^{\alpha}
\!=\!
\xi_i{\cal R}^{i \alpha}(q) ,~~
\delta \bar{q}^{\underline{\alpha }}
\!=\!
\xi_i \bar{\cal R}^{i \underline{\alpha }}(\bar{q}) ,
\label{infinitesimaltrans}
\eeq\\[-16pt]
where 
$\xi _i$
are the infinitesimal and global group parameters.
Due to the Killing equation,
the Killing vectors
${\cal R}^{i \alpha}(q)$
and
$\bar{\cal R}^{i \underline{\alpha }}(\bar{q})$
can be written locally as the gradient of some real scalar function,
the Killing potentials
${\cal M}^i (q, \bar{q})$
such that\\[-14pt]
\beq
{\cal R}^i _{~\underline{\alpha }}(q)
=
-i{\cal M}^i _{~,\underline{\alpha }} ,~~
\bar{\cal R}^i _{~\alpha}(\bar{{q}})
=
i{\cal M}^i _{~, \alpha} .
\label{gradKillingpot}
\eeq\\[-32pt]

According to van Holten et al.
\cite{NNH.01}
and using the infinitesimal $SO(2N)$ matrix $\delta g$
given by the first of
(\ref{infinitesimalop}),
the Killing potential ${\cal M}_\sigma$ 
can be written for the coset
$\frac{SO(2N)}{U(N)}$
as\\[-20pt]
\beqa
\left.
\BA{ll}
&
{\cal M}_\sigma \!
\left(
\delta a, \delta b,\delta b^\dag
\right)
\!=\!
\mbox{Tr} \!
\left( \! \delta g \widetilde{{\cal M}}_\sigma \! \right)
\!=\!
\mbox{tr}
\left(
\delta a {\cal M}_{\sigma \delta a}
\!+\!
\delta b {\cal M}_{\sigma \delta b^\dag }
\!+\!
\delta b^\dag {\cal M}_{\sigma \delta b}
\right) ,\\
\\[-14pt]
&
\widetilde{{\cal M}}_\sigma
\!\equiv\!
\left[ \!\!
\BA{cc} 
\widetilde{{\cal M}}_{\sigma \delta a} & 
\!\!\widetilde{{\cal M}}_{\sigma \delta b^\dag }\\
 \\[-8pt]
-\widetilde{{\cal M}}_{\sigma \delta b} & 
\!\!-\widetilde{{\cal M}}_{\sigma \delta a^{\!\mbox{\scriptsize T}}} 
\EA \!\!\!
\right] ,~~
\BA{c}
{\cal M}_{\sigma \delta a}
\!=\!
\widetilde{{\cal M}}_{\sigma \delta a}
\!+\!
\left( \!
\widetilde{{\cal M}}_{\sigma \delta a^{\mbox{\scriptsize T}}} \!
\right)^{\!\! \mbox{\scriptsize T}} ,\\
\\[-10pt]
{\cal M}_{\sigma \delta b}
\!=\!
\widetilde{{\cal M}}_{\sigma \delta b} ,~~
{\cal M}_{\sigma \delta b^\dag }
\!=\!
\widetilde{{\cal M}}_{\sigma \delta b^\dag } ,
\EA
\EA \!\!
\right\}
\label{KillingpotM}
\eeqa\\[-14pt]
where the trace Tr is taken over 
the $2N \!\times\! 2N$ matrices,
while the trace tr is taken over 
the $N \!\times\! N$ matrices.
Let us introduce the $N$-dimensional matrices 
${\cal R}(q; \delta g)$, 
${\cal R}_T(q; \delta g)$ and $\chi$ by\\[-22pt]
\beqa
\BA{l}
{\cal R}(q; \delta g)
\!=\!
\delta b 
\!-\! \delta a^{\mbox{\scriptsize T}} \! q \!-\! q \delta a
\!+\! q \delta b^\dag q ,~~
{\cal R}_T (q; \delta g)
\!=\!
-\delta a^{\mbox{\scriptsize T}}
\!+\! 
q \delta b^\dag ,~~
\chi
\!=\!
(1_{\!N} \!+\! q q^\dag)^{-1}
\!=\! 
\chi^\dag .
\EA
\label{RRTChi}
\eeqa\\[-22pt]
In 
(\ref{infinitesimaltrans}),
putting $\xi_i \!=\! 1$,
we have
$\delta q
\!=\! 
{\cal R}(q; \delta g)
$ 
which is the Killing vector
in the coset space
$\frac{SO(2N)}{U(N)}$,
and tr of 
holomorphic matrix-valued function 
${\cal R}_T (q; \delta g)$, namely
$\mbox{tr}[{\cal R}_T (q; \delta g)] \!=\! {\cal F}(q)$
is a holomorphic K\"{a}hler transformation. 
Then the Killing potential ${\cal M}_\sigma$ is given as\\[-20pt]
\beqa
\!\!\!\!\!\!\!\!
\left.
\BA{rl}
&-i{\cal M}_\sigma
\left(
q, \bar{q};\delta g
\right)
=
-\mbox{tr}
\Delta
\left(
q, \bar{q};\delta g
\right) ,\\
\\[-12pt]
&\Delta
\left(
q, \bar{q};\delta g
\right)
\stackrel{\mathrm{def}}{=}
{\cal R}_T (q; \delta g)
\!-\!
{\cal R}(q; \delta g) q^\dag \chi 
\!=\!
\left(
q \delta a q^\dag 
\!-\! 
\delta a^{\mbox{\scriptsize T}}
\!-\!
\delta b q^\dag 
\!+\!
q \delta b^\dag
\right) \!
\chi .
\EA \!\!
\right\}
\label{formKillingpotM} 
\eeqa\\[-10pt]
From
(\ref{KillingpotM}) and (\ref{formKillingpotM}),
we obtain\\[-16pt]
\beq
-i{\cal M}_{\sigma \delta b}
\!=\!
- \chi q ,~~
-i{\cal M}_{\sigma \delta b^\dag }
\!=\!
 q^\dag \chi ,~~
-i{\cal M}_{\sigma \delta a}
\!=\!
1_{\!N} \!-\! 2 q^\dag \chi q .
\label{componentKillingpotM} 
\eeq\\[-16pt]
Using the expression for $\widetilde{{\cal M}}_\sigma$,
equation
(\ref{componentKillingpotM}),
their components are written in the form\\[-12pt]
\beq
-i\widetilde{{\cal M}}_{\! \sigma \delta b}
\!=\!
-\chi \! q ,~
-i\widetilde{{\cal M}}_{\! \sigma \delta b^\dag }
\!=\!
 q^\dag \chi ,~
-i\widetilde{{\cal M}}_{\! \sigma \delta a}
\!=\!
-q^\dag \chi q ,~
-i\widetilde{{\cal M}}_{\! \sigma \delta a^{\mbox{\scriptsize T}}}
\!=\!
1_{\!N} \!-\! q\bar \chi q^\dag
\!=\!
\chi .
\label{tildecomponentKillingpotM} 
\eeq\\[-16pt]
It is easily checked that
the result
(\ref{componentKillingpotM})
satisfies the gradient of the real function ${\cal M}_{\sigma }$
(\ref{gradKillingpot}).
This just the Killing potential $M_\sigma$
in the $\frac{SO(2N)}{U(N)}$ coset space
obtained by van Holten et al.
\cite{NNH.01}.

To make clear the meaning of the Killing potential,
using the $2N \!\times\! N$ isometric matrix 
$u~(u^\dag u \!=\! 1_{\!N})$,
let us introduce the following 
$2N \!\times\! 2N$ matrix:\\[-10pt]
\beq
W
\!=\!
u u^\dag
\!=\!
\left[ \!\!
\BA{cc} 
R & K \\
\\[-8pt]
-\bar{K} & 1_{\!N} \!-\! \bar{R}
\EA \!\!
\right] ,
\BA{c}
R
\!=\!
b b^\dag ,\\
\\[-8pt]
K
\!=\!
b a^\dag  ,
\EA
\label{densitymat}
\eeq\\[-12pt]
which satisfies the idempotency relation 
$W^2 \!=\! W$
and is hermitian
on the $SO(2N)$ group.
The $W$ is the generalized density matrix in the $SO(2N)$ CS rep.
Since the matrices $a$ and $b$
are represented in terms of $q\!=\!(q_{\alpha\beta})$ as\\[-12pt]
\beq
a
\!=\!
(1_{\!N} \!+\! q^\dag q)^{-\frac{1}{2}} 
v ,~~
b
\!=\!
q
(1_{\!N} \!+\! q^\dag q)^{-\frac{1}{2}} 
v ,~~
v \in \! U(N) ,
\label{matAandB}
\eeq\\[-18pt]
then, we have\\[-20pt]
\beq
R
\!=\!
q (1_{\!N} \!+\! q^\dag q)^{-1} q^\dag 
\!=\!
q \bar{\chi }q^\dag
\!=\!
1_{\!N} \!-\! \chi ,~~
K
\!=\!
q
(1_{\!N} \!+\! q^\dag q)^{-1}
\!=\!
\chi q .
\label{matRandK}
\eeq\\[-16pt]
To our great surprise,
substituting
(\ref{matRandK})
into
(\ref{tildecomponentKillingpotM}),
the expression for the Killing potential
$-i\overline{\widetilde{{\cal M}}}_\sigma$
just becomes equivalent with
the generalized density matrix
(\ref{densitymat}).
This fact first has been found by the present authors in Ref.
\cite{SJCF.08}.
The two relations in
(\ref{matAandB})
play an important role to make another way of the derivation of the TDRHB equation
as shown in Appendix C.

First according to
\cite{NNH.01}
we define a matrix
$\Xi (q)$ and require a transformation rule as follows:\\[-10pt]
\beq
\Xi (q)
\!\stackrel{\mathrm{def}}{=}\!
\left[ \!\!\!
\BA{cc}
1_{\!N} & 0 \\
\\[-16pt]
q & 1_{\!N} \\
\EA \!\!\!
\right] ,~
\Xi^{-1} (q)
\!=\!
\Xi (-q),
\label{matXi}
\eeq
\vspace{-0.9cm}
\beqa
\!\!\!\!
\Xi (q)
\!\longrightarrow\!
\Xi ({}^g \! q)
\!=\!
g \Xi (q) \widehat{H}^{-1}(q; g),
~\mbox{with}~
\widehat{H} (q; g)
\!=\!
\left[ \!\!\!
\BA{cc}
\left( \! \widehat{H}_+ (q; g) \! \right)^{\!\!-1} &\!\!\!\!
\widehat{H}_0 (q; g) \\
\\[-12pt]
0 &\!\!\!\! \widehat{H}_- (q; g)
\EA \!\!\!
\right] \! .
\label{transmatXi}
\eeqa\\[-12pt]
It must be remarkable that
as is clear from the structure of the transformation
(\ref{transmatXi}),
in the above transformation 
a constant unitary matrix $g$ can be canceled by taking a bilinear form 
$\Xi ^{\dag}({}^g \! q) \Xi ({}^g \! q)$
\cite{BKMU.84}.
The
$
{}^g \! q
$
satisfying
$
\left( {}^g \! q \right)^{\!\mbox{\scriptsize T}}
\!\!=\!
{}^g \! q^{\mbox{\scriptsize T}}
\!\!=\!
-{}^g \! q
$
is a nonlinear M\"{o}bius transformation
given by\\[-22pt]
\beqa
\!\!\!\!
{}^g \! q
\!=\!
(b \!+\! \bar{a} q)
(a \!+\! \bar{b} q)^{-1}
\!=\!
-
(a^{\mbox{\scriptsize T}} \!-\! q b^\dag )^{-1}
(b^{\mbox{\scriptsize T}} \!-\! q a^\dag ) ,
\label{MtransmatQ}
\eeqa\\[-24pt]
which obeys a successive transformation rule 
$
{}^{g^\prime } \! (\!{}^g \! q \!)
\!=\!
{}^{g^\prime \! g} \! q
$,
$\!$i.e., composition of two transformations
$g^\prime$ and $g$.
The above nonlinear M\"{o}bius transformation makes a crucial role
to construct a solution of the TDRHB equation
in the proceeding sections.
Under an action of $SO(2N)$ matrix $g$, i.e.,
last equation of
(\ref{Bogotrans}),
using
$g^{-1} \!\!=\!\! g^{\dag }$,
(\ref{transmatXi})
and
(\ref{MtransmatQ}),
we have the relation
$\widehat{H} (q; g)
 \!=\!
\Xi (-{}^g \! q) ~\! g ~\! \Xi (q)
$.
Then
$\widehat{H} (q; g)$
takes a form\\[-8pt]
\beq
\widehat{H} (q; g)
\!=\!\!
\left[ \!\!\!
\BA{cc}
a \!+\! \bar{b} q &\!\!\! \bar{b} \\
\\[-4pt]
0 &\!\!\! (a^{\mbox{\scriptsize T}}
\!-\!
q b^\dag )^{\!-1}
\EA \!\!\!
\right] \! ,~
\widehat{H}_+ (q; g)
\!=\!
\widehat{H}_- ^{\mbox{\scriptsize T}} (q; g) ,~
\det \widehat{H}_+ (q; g)
\!=\!
\det \widehat{H}_- (q; g) .
\label{hatHform}
\eeq\\[-12pt]
Multiplying $g^\prime$ by $g$,
we also have useful product formulas for
$\widehat{H}_{\pm,0} (q; g^\prime g)$
as\\[-20pt]
\beqa
\!\!\!\!\!\!\!\!\!
\BA{c}
\widehat{H}_{\!+} \! (q; \! g^\prime \! g)
\!=\!
\widehat{H}_{\!+} \! (q; \! g) \!
\widehat{H}_{\!+} \! ({}^g \! q; \! g^\prime), 
\widehat{H}_{\!-} \! (q; \! g^\prime \! g)
\!=\!
\widehat{H}_{\!-} \! ({}^g \! q; \! g^\prime) \!
\widehat{H}_{\!-} \! (q; \! g) ,
\widehat{H}_{\!0} (q; \! g^\prime \! g)
\!=\!
\widehat{H}_{\!0} ({}^g \! q; \! g^\prime) \!
\widehat{H}_{\!0} (q; \! g)
\!=\!
\bar{b^\prime}\bar{b} .
\EA \!\!\!\!
\label{decompoHminus}
\eeqa\\[-20pt]
Then we have the product formula
$\widehat{H} (q; \! g^\prime g)
\!=\!
\widehat{H} ({}^g \! q; \! g^\prime) 
\widehat{H} (q; \! g)$.

Using the matrices $a$ and $b$ given by
(\ref{matAandB})$(v \!=\! 1_{\!N})$,
we obtain the matrix decomposition\\[-10pt]
\beq
\!\!
g
\!=\!
\Xi ^{\dag}(-q) \!\!
\left[ \!\!\!\!
\BA{cc}
(1_{\!N} \!\!+\!\! q^\dag \! q)^{\frac{1}{2}}  &\!\!\!\!\!\!\! 0 \\
\\[-6pt]
0 &\!\!\!\!\!\!\! (1_{\!N} \!\!+\!\! q q^\dag \! )^{\!-\frac{1}{2}} 
\EA \!\!\!\!
\right] \!\!
\Xi (q)
\!=\!
\Xi ^{\dag} \!\!
\left( \!\!
- 
{\displaystyle \lambda \frac{\tan(\! \sqrt{\! \lambda^\dag \! \lambda})}
{\sqrt{\!\lambda^\dag \! \lambda}}} \!\!
\right) \!\!\!
\left[ \!\!\!
\BA{cc}
\cos^{\!-1} \! (\! \sqrt{\!\lambda^\dag\!  \lambda})  &\!\!\!\!\!\!\! 0 \\
\\[-6pt]
0 &\!\!\!\!\!\!\! \cos(\! \sqrt{\! \lambda^\dag \! \lambda}) 
\EA \!\!\!\!
\right] \!\!
\Xi \!
\left( \!\!
\lambda
{\displaystyle \frac{\tan(\! \sqrt{\! \lambda^\dag \! \lambda})}{\sqrt{\! \lambda^\dag \! \lambda}}} \!\!
\right) \! ,
\label{decomposition}
\eeq\\[-14pt]
whose relation substituted by the $q$
(\ref{Bogocoset})
is the Harish-Chandra decomposition.
Such a relation is derived based on the nonlinear M\"{o}bius transformation
(\ref{MtransmatQ}).
It should be noticed that the geodesic flow through the identity coset element
corresponds to
$
\lambda 
{\displaystyle \frac{\tan(\! t \sqrt{\! \lambda^\dag \! \lambda})}{\sqrt{\! \lambda^\dag \! \lambda}}}
$.
These facts have already been pointed out by Berceanu et. al.
\cite{BG.91}.
Finally, from the second equation of
(\ref{decomposition}),
a $SO(2N)$ matrix $g$ is expressed in terns of the original variable $\lambda$
contained in the generator $\Lambda$.
For the sake of convenience
we redefine the K\"{a}hler potential as~
$
{\cal K}(q,\overline{q})
\!=\!
\ln \det
\left(
1_{\!N}
\!+\!
q \overline{q}
\right)
$.
Under the nonlinear M\"{o}bius transformation
(\ref{MtransmatQ}),
the K\"{a}hler potential
transforms as~
$
{\cal K} ({}^g \! q,
{}^g \overline{q})
\!=\!
{\cal K}(q ,\overline{q})
\!+\!
{\cal F}(q ;g)
\!+\!
\overline{{\cal F}}
(\overline{q};g) 
$.

Next according to
\cite{NNH.01},
to require anomaly cancellations with matter fields,
one may change an  assignment of $U(1)$ charges
by introducing a complex line bundle ${\cal S}$
which is defined as a complex matter scalar field
coupled to the SUSY $\sigma$-model
with the infinitesimal transformation law
$
\delta_{i} {\cal S}^\lambda
\!\!=\!\!
\lambda {\cal F}_{i} (q) {\cal S}
$.
For a tensor rep 
$
{\cal T}^{\alpha_1 \cdots \alpha_p}
\!\!\equiv\!\!
{\cal S}^\lambda T^{\alpha_1 \cdots \alpha_p}
$,
${\cal T}$
obeys the transformation rule\\[-22pt]
\beqa
\begin{array}{c}
\delta_{i} {\cal T}^{\alpha_1 \cdots \alpha_p}
\!=\!
\sum_{k \!=\! 1} ^p
{\cal R}^{\alpha_k}_{{i},~\beta }(q)
{\cal T}^{\alpha_1 \cdots \beta \dots \alpha_p}
\!+\!
\lambda {\cal F}_{i} (q)
{\cal T}^{\alpha_1 \cdots \alpha_p} .
\end{array}
\label{infinitesimaltransT}
\eeqa\\[-24pt]
A section of a minimal line bundle over
$\frac{SO(2N)}{U(N)}$
is given by
$
{}^g {\cal S}
\!\!=\!\!
\left[ \!
\det \! \widehat{H}_+ (q; g) \!
\right]^{\!\frac{1}{2}} \!\! {\cal S}
\!\!=\!\!
\left[ \!
\det \! \widehat{H}_- (q; g) \!
\right]^{\!\frac{1}{2}} \!\! {\cal S}
$.
Suppose that
${\cal T}_{(p;q)}^{i_1 \!\cdots\! i_p}$
is an irreducible completely antisymmetric $SU(N)$-tensor
rep with $p$ and $q$ indices.
We abbreviate it simply as ${\cal T}_{(p;q)}$.
By taking the completely antisymmetric tensor product of
$SU(N)$ vectors
${\cal T}_1 ^{i_1}, \!\cdots\!, {\cal T}_p ^{i_p}$,
we obtain a $SU(N)$ tensor of rank $p$
with index $q$
as\\[-12pt]
\beq
{\cal T}_{(p;q)} ^{i_1 \cdots i_p}
\!\equiv\!
\frac{1}{p!}
{\cal S}^q T_1 ^{[ i_1} \ast \cdots \ast T_p ^{i_p ]} ,
\label{SUNplus1tensor}
\eeq\\[-14pt]
where
$[ \cdots ]$
denotes the completely anti-symmetrization of the indices
inside the brackets.
Thus we obtain a transformation of tensor
${\cal T}_{(p;q)} ^{i_1 \cdots i_p}$
as\\[-12pt]
\beq
{}^g{\cal T}_{(p;q)} ^{i_1 \cdots i_p}
\!=\!
\left[
\det \widehat{H}_- (q; g)
\right]^{\!\frac{q}{2}} \!
\left[
\widehat{H}_- (q; g)
\right]_{~~j_1} ^{i_1} \!
\cdots \!
\left[
\widehat{H}_- (q; g)
\right]_{~~j_p} ^{i_p} \!
{\cal T}_{(p;q)} ^{j_1 \cdots j_p} .
\label{transSUNplus1tensor}
\eeq\\[-10pt]
The invariant K\"{a}hler potential for a tensor
is given by\\[-20pt]
\beqa
{\cal K}_{(p;q)}
\!=\!
\overline{{\cal T}}_{(p;q) j_1 \cdots j_p}
{\cal G}_{(p;q) i_1 \cdots i_p} ^{j_1 \cdots j_p}
{\cal T}_{(p;q)} ^{i_1 \cdots i_p} ,~
{\cal G}_{(p;q) i_1 \cdots i_p} ^{j_1 \cdots j_p}
\!\equiv\!
\frac{1}{p!}
[\det {\chi}]^{\frac{q}{2}}
{\chi}_{~~i_1} ^{j_1}
\cdots
{\chi}_{~~i_p} ^{j_p} .
\label{KpotSUNplus1tensor}
\eeqa\\[-14pt]
A $SU(N)$ dual tensor
$
{\cal T}_{(\overline{N \!-\! p};q)
i_{p \!+\! 1} \cdots i_{N}}
$
with $( N \!-\! p )$ indices and
index $q$ is\\[-14pt]
\beqa
{\cal T}_{(\overline{N \!-\! p};q)
i_{p \!+\! 1} \cdots i_{N1}}
\!\equiv\!
\frac{1}{p!}
{\cal T}_{(p;q)} ^{i_p \cdots i_1}
\epsilon_{i_1 \cdots i_{N}} ,~
(\epsilon_{i_1 \cdots i_{N}}: SU(N)
\mbox{Levi-Civita tensor} )
\label{dualSUNplus1tensor}
\eeqa\\[-12pt]
which transforms under the nonlinear M\"{o}bius transformation
(\ref{MtransmatQ})
as\\[-18pt]
\beqa
{}^g{\cal T}_{(\overline{p};q) i_1 \cdots i_p}
\!=\!
{\cal T}_{(\overline{p};q) j_1 \cdots j_p}
\left[
\widehat{H}_- ^{-1} (q; g)
\right]_{~~i_1} ^{j_1}
\cdots
\left[
\widehat{H}_- ^{-1} (q; g)
\right]_{~~i_p} ^{j_p} \!
\left[
\det \widehat{H}_- (q; g)
\right]^{1 \!+\! \frac{q}{2}} .
\label{transdualSUNplus1tensor}
\eeqa\\[-16pt]
The invariant K\"{a}hler potential for a dual tensor
is given by\\[-16pt]
\beqa
\begin{array}{l}
{\cal K}_{(\overline{p};q)}
\!=\!
{\cal T}_{(\overline{p};q) i_1 \cdots i_p}
{\cal G}_{(\overline{p};q) j_1 \cdots j_p} ^{i_1 \cdots i_p}
\overline{{\cal T}}_{(\overline{p};q)} ^{j_1 \cdots j_p} ,~
{\cal G}_{(\overline{p};q) j_1 \cdots j_p} ^{i_1 \cdots i_p}
\!\equiv\!
{\displaystyle \frac{1}{p!}}
[\det {\chi}]^{1 \!+\! \frac{q}{2}} \!
\left[ {\chi}^{-1} \right]_{~~j_1} ^{i_1}
\cdots
\left[ {\chi}^{-1} \right]_{~~j_p} ^{i_p}.
\end{array}
\label{KpotSUNplus1dualtensor}
\eeqa\\[-26pt]

The contributions of the invariant K\"{a}hler potentials
${\cal K}_{(p;q)}$
and
${\cal K}_{(\overline{p};q)}$
to the Killing potentials,
${\cal M}_{(p;q)}(q,\overline{q}; \delta g )$
and
${\cal M}_{(\overline{p};q)}(q,\overline{q}; \delta g )$
for a tensor
${\cal T}_{(\overline{p};q)}$
and a dual tensor
$\overline{{\cal T}}_{(\overline{p};q)}$
of rank $p$ with index $q$,
are obtained
to satisfy
${\cal F}_i (q) \!=\! 0$
and
$\overline{{\cal F}}_i ( \overline{q}) \!=\! 0$
as\\[-16pt]
\beqa
\begin{array}{c}
-i {\cal M}_{\left( \! \binom{p}{\overline{p}};q \! \right)} 
\left(\! q,\overline{q}; \delta g \! \right)
=
{\cal K}_{\left( \! \binom{p}{\overline{p}};q \! \right),~[\alpha]}
\left( \! q,\overline{q}; \delta g \! \right)
{\cal R}^{[\alpha]} (q) .
\end{array}
\label{KllingpotM}
\eeqa\\[-16pt]
From
(\ref{KllingpotM})
and
(\ref{KpotSUNplus1tensor}),
the Killing potential for tensors
is exactly computed as\\[-16pt]
\beqa
\!\!\!\!
\begin{array}{rl}
- i {\cal M}_{(p;q)}
\!\!\!&\!=\!
{\cal K}_{(p;q),~[i]} {\cal R}^{[i]}
\!=\!
\overline{{\cal T}}_{(p;q) j_1 \cdots j_p,~[i]} {\cal R}^{[i]}
{\cal G}_{(p;q) i_1 \cdots i_p} ^{j_1 \cdots j_p}
{\cal T}_{(p;q)} ^{i_1 \cdots i_p} \\
\\[-12pt]
&\!+\!
\overline{{\cal T}}_{(p;q) j_1 \cdots j_p}
{\cal G}_{(p;q) i_1 \cdots i_p,~[i]} ^{j_1 \cdots j_p}
{\cal R}^{[i]}
{\cal T}_{(p;q)} ^{i_1 \cdots i_p}
\!+\!
\overline{{\cal T}}_{(p;q) j_1 \cdots j_p}
{\cal G}_{(p;q) i_1 \cdots i_p} ^{j_1 \cdots j_p}
{\cal T}_{(p;q),~[i]} ^{i_1 \cdots i_p}
{\cal R}^{[i]} ,~([i]\!=\!(i\hat{i})) ,
\end{array}
\label{variKpotSUNplus1tensor}
\eeqa\\[-6pt]
Due to our recent work
\cite{SJCF.11}, 
the variation of
$\delta {\cal G}_{(p;q) i_1 \cdots i_p} ^{j_1 \cdots j_p}$
is calculated as\\[-16pt]
\beqa
\begin{array}{rl}
\delta {\cal G}_{(p;q) i_1 \cdots i_p} ^{j_1 \cdots j_p}
\!\!\!&=
-
{\displaystyle \frac{q}{2} \frac{1}{p!}}
[\det {\chi}]^{\frac{q}{2}}
\mbox{tr} \!
\left\{ \!
{\chi} \!
\left(
\delta q \overline{q}
\!+\!
q \delta \overline{q}
\right) 
\right\}
{\chi}_{~~i_1} ^{j_1}
\cdots
{\chi}_{~~i_p} ^{j_p} \\
\\[-12pt]
&~~
-
{\displaystyle \frac{1}{p!}}
[\det {\chi}]^{\frac{q}{2}} \!
\sum_{r\!=\!1} ^p
{\chi}_{~~i_1} ^{j_1}
\cdots
\left\{
{\chi} \!
\left(
\delta q \overline{q}
\!+\!
q \delta \overline{q}
\right) 
{\chi} \!
\right\}_{~~i_r} ^{j_r}
\cdots
{\chi}_{~~i_p} ^{j_p} ,
\end{array}
\label{delGSUNplus1tensor}
\eeqa\\[-12pt]
together with the variations\\[-16pt]
\beq
\delta \det \! {\chi}
\!\!=\!
- \det {\chi}
\!\cdot\!
\mbox{tr} \!
\left\{
{\chi} \!
\left(
\delta q \overline{q}
\!+\!
q \delta \overline{q}
\right) 
\right\} ,~
\delta {\chi}_{~i}^j
\!=\!
-
\left\{
{\chi} \!
\left(
\delta q \overline{q}
\!+\!
q \delta \overline{q}
\right) \!
{\chi}
\right\}_{~i}^j .
\eeq\\[-16pt]
Taking only the
$\delta q$
term
in
(\ref{delGSUNplus1tensor}),
the following type of contraction is easily carried out:\\[-18pt]
\beqa
\!\!\!\!
\begin{array}{rl}
{\cal G}_{(p;q) i_1 \cdots i_p,~\hat{i}i} ^{j_1 \cdots j_p}
\delta {q}_{i\hat{i}}
\!\!\!&=\!
- {\displaystyle \frac{q}{2}\frac{1}{p!}}
[\det {\chi}]^{\frac{q}{2}}
\mbox{tr} \left( {\cal R}_T \!-\! \Delta \right)
{\chi}_{~~i_1} ^{j_1}
\cdots
{\chi}_{~~i_p} ^{j_p} \\
\\[-14pt]
&~~
-
{\displaystyle \frac{1}{p!}}
[\det {\chi}]^{\frac{q}{2}} \!
\sum_{r\!=\!1} ^p
{\chi}_{~~i_1} ^{j_1}
\cdots
\left\{ \!
{\chi}_{~~i} ^{j_r}
\left( {\cal R}_T \!-\! \Delta \right)_{~i_r} ^{i} \!
\right\}
\cdots
{\chi}_{~~i_p} ^{j_p} ,
\end{array}
\label{sumdelGSUNplus1tensor}
\eeqa\\[-10pt]
where we have used the relation
$
\delta q \overline{q} {\chi}
\!=\!
{\cal R}_T \!-\! \Delta  
$.
The $(i\hat{i})$ element of the matrix
${q}$, i.e.,
${q}_{i\hat{i}}$
denoted as ${q}^{[i]}$,
$(\hat{i}\mbox{: another component different from}~i)$ and
${\cal R}^{[i]}$ are given by the Killing vector, i.e.,
$
{\cal R}^{[i]}
\!=\!
\delta q^{[i]}
$,
which is the first equation of
(\ref{infinitesimaltrans})
with
$\xi_l \!=\! 1$.
The equation
$
\overline{{\cal T}}_{(p;q) j_1 \cdots j_p,[i]} {\cal R}^{[i]}
\!=\!
0
$
is evident.
Similarly, the Killing potential for dual tensors is also computed.
Then we reach\\[-16pt]
\beqa
\!\!\!\!\!\!\!\!
\left.
\BA{ll}
&
\begin{array}{l}
-i {\cal M}_{(p;q)} (\! {q},\overline{q}; \delta {\cal G} )
\!=\!
{\displaystyle \frac{1}{p!}}
\overline{{\cal T}}_{(p;q) j_1 \cdots j_p} \\
\\[-12pt]
~~
\times
[{\det {\chi}}]^{\frac{q}{2}}
{\chi}_{~~k_1} ^{j_1}
\cdots
{\chi}_{~~k_p} ^{j_p} \\
\\[-10pt]
~~
\times 
\left\{ \!
\sum_{r \!=\! 1} ^p
\delta_{~~i_1} ^{k_1}
\cdots
\left[ 
\Delta(\! {q},\overline{q};
\delta {\cal G} ) 
\right]
_{~~i_r} ^{k_r}
\cdots
\delta_{~~i_p} ^{k_p}
\!+\!
{\displaystyle \frac{q}{2}}
\mbox{tr}
\left[ 
\Delta ( {q},\overline{q}; \delta {\cal G} )
\right]
\delta_{~~i_1} ^{k_1}
\cdots
\delta_{~~i_p} ^{k_p} \!
\right\} \!
{\cal T}_{(p;q)} ^{i_1 \cdots i_p} ,
\end{array} \\
\\[-12pt]
&
\begin{array}{l}
-i {\cal M}_{(\overline{p};q)} (\! {q},\overline{q}; \delta {\cal G} )
\!=\!
{\displaystyle \frac{1}{p!}}
{\cal T}_{(\overline{p};q) j_1 \cdots j_p} \\
\\[-12pt]
~~
\times
\left\{ \!
\sum_{r \!=\! 1} ^p
\delta_{~~k_1} ^{j_1}
\!\cdots\!
[ -\Delta (\! {q},\overline{q};
\delta {\cal G} ) ]_{~~k_r} ^{j_r}
\!\cdots\!
\delta_{~~k_p} ^{j_p}
\!+\!
\left(
{\displaystyle \! 1 \!+\! \frac{q}{2}}
\right)
\mbox{tr}
\left[
\Delta (\! {q},\overline{q}; \delta {\cal G} )
\right]
\delta_{~~k_1} ^{j_1}
\!\cdots\!
\delta_{~~k_p} ^{j_p} \!
\right\} \\
\\[-10pt]
~~
\times 
[{\det {\chi}}]^{1 \!+\! \frac{q}{2}}
[{\chi}^{-1}]_{~~i_1} ^{k_1}
\cdots
[{\chi}^{-1}]_{~~i_p} ^{k_p}
\overline{{\cal T}}_{(\overline{p};q)} ^{i_1 \cdots i_p} ,
\end{array}
\EA \!\!
\right\}
\label{KllingpotM}
\eeqa
which is useful to optimize the Killing potential.
$\!\!$The $\lambda$ is a power of complex line bundle $\!{\cal S}\!$,
${\cal S}^{\!\lambda}$
given by
$\lambda \!\!=\!\! {\displaystyle \frac{\bf q}{2}}$
in the upper
and 
$\lambda \!\!=\!\! 1 \!\!+\!\! {\displaystyle \frac{\bf q}{2}}$
in the lower.
The $\bf q$ stands for a rescaling charge.


\newpage

\setcounter{equation}{0}
\renewcommand{\theequation}{\arabic{section}.\arabic{equation}}

\section{Two-level model}

~~~
In this Section, we use a spherical symmetric single-particle state specified by
the set of quantum numbers $\{n_a,l_a,j_a,m_\alpha\}$,
denoted as $\alpha~(\alpha \! =\! 1, \cdots, N)$.
The time-reversed single-particle state
$\bar\alpha$ is obtained from $\alpha$ by changing the sign of
$m_\alpha$.
We use a phase factor $s_\alpha \!=\! (-1)^{j_a-m_\alpha }$ in the time-reversed quantity.
The contribution of the pair interaction 
to the Hartree-Fock (HF) potential
$[\alpha\beta|\gamma\delta] R_{\gamma\delta}$
is neglected.
Further assume the pairing potential
$D\!=\!(D_{\alpha \beta })$
to be constant.
This makes the situation very simple as in the BCS theory does
\cite{RS.80}.
Then the $SO(2N)$ SCF parameters
$F_{\alpha\beta }(t)$ and $D_{\alpha\beta }(t)$
defined by
(\ref{SCFparameters})
 have the following forms:\\[-16pt]
\ba
\left.
\BA{l}
F_{\alpha\beta }(t)
\!=\!
(\varepsilon_a \!-\! \lambda) \!\cdot\! \delta_{\alpha\beta } ,\\
\\[-14pt]
D_{\alpha\beta }(t)
\!=\!
- s_\alpha \delta_{\alpha \bar\beta }\Delta(t) ,~
\Delta(t)
\!\equiv\!
{\displaystyle \frac{1}{2}} g \!
\sum_{\gamma} \!
s_\gamma K_{\gamma \bar\gamma }(t) ,~
s_\gamma K_{\gamma \bar\gamma }(t)
\!=\!
s_\gamma \!
\left[ q (1_{\!N} \!-\!  \bar{q} q)^{-1} \right]_{\gamma \bar\gamma } ,
\EA \!\!
\right\}
\label{usualSCFparameters}
\ea\\[-12pt]
where $g$ is the strength parameter for the pairing force.

First we treat a simple twe-level model $(N \!=\! 2)$
of a fermion system under consideration.
We denote the quantities
$q_{12}(t)$ simply as $q(t)$, 
$(\! \varepsilon_1 \!-\! \lambda \!) \!+\! (\! \varepsilon_2 \!-\! \lambda \!)$ as
$2 \varepsilon$
and
$\!D_{12}(t)$ as\\[-18pt]
\ba
\left.
\BA{l}
-{\cal D}(t) \!=\! s_1 \delta_{1 \bar2 }\Delta(t),
~
\Delta(t)
\!\equiv\!
{\displaystyle \frac{1}{2}} g \!
\left(
s_1 K_{1 \bar1 }(t)
\!+\!
s_2 K_{2 \bar2 }(t)
\right) , \\
\\[-12pt]
s_1 K_{1 \bar1 }(t)
\!=\!
s_1 \!\! 
\left[ \!
\left[ \!\!
\BA{cc}
0 & q \\
\\[-14pt]
-q & 0 
\EA \!\!
\right] \!\!
\left( \!
\left[ \!\!
\BA{cc}
1 & 0 \\
\\[-14pt]
0 & 1 
\EA \!\!
\right]
\!\!-\!\!
\left[ \!\!
\BA{cc}
0 & \bar{q} \\
\\[-14pt]
-\bar{q} & 0
\EA \!\!
\right] \!\!
\left[ \!\!
\BA{cc}
0 & q \\
\\[-14pt]
-q & 0 
\EA \!\!
\right] \!
\right)^{\!\!-1}
\right]_{\!1 \bar1}
\!\!\!=\!
s_1
{\displaystyle \frac{1}{1 \!\!+\!\! |q|^{2}}} \!
\left[ \!
\BA{cc}
0 & q \\
\\[-14pt]
-q & 0 
\EA \!
\right]_{\!1 \bar1} \\
\\[-12pt]
~~~~~~~~~~~  
\!=\!
~s_1 
{\displaystyle \frac{1}{1 \!\!+\!\! |q|^{2}}} q ,~
(\mbox{matrix element:}~1 \bar1 \!=\! 12) , \\
\\[-12pt]
s_2 K_{2 \bar2 }(t)
\!=\!
s_2 
{\displaystyle \frac{1}{1 \!\!+\!\! |q|^{2}}} \! 
\left[ \!
\BA{cc}
0 & q \\
\\[-14pt]
-q & 0 
\EA \!
\right]_{\!2 \bar2} 
\!=\!
- 
s_2 
{\displaystyle \frac{1}{1 \!\!+\!\! |q|^{2}}} q , ~\!
(\mbox{matrix element:}~2 \bar2 \!=\! 21) , \\
\\[-8pt]
\EA \!\!\!
\right\}
\label{SCFdeltas}
\ea\\[-28pt]

Then we have the following Ricatti equations:\\[-16pt]
\ba
\BA{ll}
\left[ \!\!
\BA{c}
 \dot q(t) \\
\\[6pt]
\dot{\bar{q}}(t) 
\EA \!\!
\right]
\!=\!\!
\left[ \!\!
\BA{cc}
{\displaystyle
\frac{{\cal D}(t)}{i\hbar } 
\!+\!
2 \frac{\varepsilon}{i\hbar } q(t)
\!+\!
\frac{\bar{\cal D} (t) }{i\hbar }q(t)^2 
}\\
\\[-10pt]
{\displaystyle
-\frac{\bar{{\cal D}}(t) }{i\hbar }
\!-\!
2 \frac{\varepsilon}{i\hbar } \bar{q}(t)
\!-\! \frac{{\cal  D}(t) }{i\hbar } \bar{q}(t)^2 
}
\EA \!\!
\right] \! ,
\EA
\label{RiccatiEqN2}
\ea
and define the matrix $A(t)$ as\\[-16pt]
\ba
A(t)
\!\equiv\!
\left[ \!\!
\BA{cc}
{\displaystyle \frac{\varepsilon}{i\hbar }} & {\displaystyle \frac{{\cal D}(t)}{i\hbar }} \\
\\[-10pt]
-{\displaystyle \frac{\bar{{\cal D}}(t) }{i\hbar }} & -{\displaystyle \frac{\varepsilon}{i\hbar }} 
\EA \!\!
\right] \! ,
\BA{c}
\det  \! A(t)
\!=\!
{\displaystyle
\frac{\varepsilon^2}{\hbar^2}
\!-\!
\frac{|D(t)|^2}{\hbar^2} 
}
\!\equiv\!
\delta^2 , \\
\\[-2pt]
A(t)^2
\!=\!
\left(- \det  \! A(t)\right)
\!\cdot\!
1_2 .
\EA 
\label{RiccatiMat}
\ea
We give a solution for $q(t)$ up to the first order and the infinite order in $t$,
respectively, as\\[-18pt]
\ba
\exp\{tA(t)\}
\!\!=\!\!
1_2
\!+\!
t
\!\cdot\!
A(t)
\left( \delta^2 \!=\! 0 \right), ~
\exp\{tA(t)\}
\!\!=\!\!
\BA{c}
\cos \! \left( t \delta \right)  \\
\\[-8pt]
\cosh \! \left( t \delta \right) 
\EA \!\!\!
\!\cdot\!
1_2
\!+\!
{\displaystyle \frac{1}{\delta}} \!\!
\BA{c}
\sin \! \left( t \delta \right)  \\
\\[-8pt]
\sinh \! \left( t \delta \right) 
\EA \!\!\!
\!\cdot\!
A(t) 
\BA{c}
\left( \delta^2 \!>\! 0 \right)  \\
\\[-10pt]
\left( - \delta^2 \!<\! 0 \right)
\EA \!\!\!.
\label{expRiccatiMat}
\ea\\[-12pt]
Following Inoguchi
\cite{Inoguchi.01},
the above solution
is shown to satisfy the Riccati equation
as follows:\\[-14pt]
\ba
\BA{ll}
\left. {\displaystyle \frac{d}{dt }} \! \right|_{t \!=\! 0} \!\!
q \! \left( \! T_{\exp\{ \! tA(t)\} \! } \! (\!q\!) \! \right)
\!\!=\!\!
\left. {\displaystyle \frac{d}{dt }} \! \right|_{t \!=\! 0} \!\!
{\displaystyle
\frac{
\BA{c}
\cos \! \left( \! t \delta \! \right)
\!\!+\!\!
{\displaystyle \frac{1}{\delta}} \!
\sin \! \left( \! t \delta \! \right) \!
{\displaystyle \frac{\varepsilon}{i\hbar }} \\
\\[-12pt]
\cosh \! \left( \! t \delta \! \right) 
\!\!+\!\!
{\displaystyle \frac{1}{\delta}} \!
\sinh \! \left( \! t \delta \! \right) \!
{\displaystyle \frac{\varepsilon}{i\hbar }} 
\EA \!\!\!
\!\!\cdot\!
q
\!+\!
{\displaystyle \frac{1}{\delta}} \!\!\!
\BA{c}
\sin \! \left( \! t \delta \! \right) \!
{\displaystyle \frac{{\cal D}(t)}{i\hbar }} \\
\\[-10pt]
\sinh \! \left( \! t \delta \! \right) \!
{\displaystyle \frac{{\cal D}(t)}{i\hbar }} 
\EA \!\!\! 
}
{
\BA{c}
\cos \! \left( \! t \delta \! \right)
\!-\!
{\displaystyle \frac{1}{\delta}} \!
\sin \! \left( \! t \delta \! \right) \!
{\displaystyle \frac{\varepsilon}{i\hbar }}  \\
\\[-12pt]
\cosh \! \left( \! t \delta \! \right)
\!-\!
{\displaystyle \frac{1}{\delta}} \!
\sin \! \left( \! t \delta \! \right) \!
{\displaystyle \frac{\varepsilon}{i\hbar }} 
\EA \!\!\!
-
{\displaystyle \frac{1}{\delta}} \!\!\!
\BA{c}
\sin \! \left( \! t \delta \! \right) \!
{\displaystyle \frac{\bar{{\cal D}}(t) }{i\hbar }}  \\
\\[-10pt]
\sinh \! \left( \! t \delta \! \right) \!
{\displaystyle \frac{\bar{{\cal D}}(t) }{i\hbar }}
\EA \!\!\!
\!\!\cdot\!
q 
}
}
\!\!=\!\!
{\displaystyle
\frac{{\cal D}(t)}{i\hbar } 
\!\!+\!\!
2 \! \frac{\varepsilon}{i\hbar } \! q(t)
\!\!+\!\!
\frac{\bar{\cal D} (t) }{i\hbar } \! q(t)^2 
} \! ,
\EA
\label{RiccatiSol}
\ea\\[-12pt]
from which,
the Riccati equation for $q(t)$
(\ref{RiccatiEqN2})
can be surely derived.
The case $\delta^2 \!\!=\!\! 0$ is trivially derived.
In Appendix D,
we treat three- and four-level models $(N \!=\! 3 ~\mbox{and}~4)$.


\newpage

\setcounter{equation}{0}
\renewcommand{\theequation}{\arabic{section}.\arabic{equation}}

\section{General solution of ${\bf \frac{SO(2N)}{U(N)}}$
TDRHB equation}

~~
Following the way adopted in the preceding Section,
we construct a general
solution of time dependent ${\displaystyle \frac{SO(2N)}{U(N)}}$
Riccati-Hartree-Bogoliubov (TDRHB) equation.
We here use a modified matrix $\frac{{\cal F}}{i\hbar }$ together with 
$N \!\times\! N$ matrices $\delta^2$ and $\bar{\delta}^2$ and suppose relations\\[-16pt]
\ba
{\displaystyle \frac{{\cal F}}{i\hbar }}
\!\equiv\!
\left[ \!\!
\BA{cc}
{\displaystyle \frac{F}{i\hbar }} & {\displaystyle \frac{D}{i\hbar }} \\
\\[-10pt]
-{\displaystyle \frac{\bar{D}}{i\hbar }} & -{\displaystyle \frac{\bar{F}}{i\hbar }} 
\EA \!\!
\right] \! ,
\BA{c}
- \delta^2
\!\equiv\!
{\displaystyle
\frac{F^2}{(i\hbar)^2}
\!-\!
\frac{D\bar{D}}{(i\hbar)^2} 
} ,~
FD
\! =\!
D\bar{F} , \\
\\[-10pt]
- \bar{\delta}^2
\!\equiv\!
{\displaystyle
\frac{\bar{F}^2}{(i\hbar)^2}
\!-\!
\frac{\bar{D}D}{(i\hbar)^2} 
},~
\bar{F}\bar{D}
\! =\!
\bar{D}F .
\EA 
\label{RiccatiMat2}
\ea\\[-10pt]
We give a solution for $q(t)$ up to the first order and the infinite order in $t$,
respectively, as\\[-16pt]
\ba
\exp\{t{\displaystyle \frac{{\cal F}}{i\hbar }}\}
\!\!=\!\!
1_{2N}
\!+\!
t
\!\cdot\!
{\displaystyle \frac{{\cal F}}{i\hbar }}
\left( ||\delta^2|| \!=\! 0 \right), 
\label{expRiccatiMat2}
\ea
\vspace{-0.4cm}
\ba
\exp\{t{\displaystyle \frac{{\cal F}}{i\hbar }}\}
\!\!=\!\!
\left[ \!\!\!
\BA{cc}
\cos \! \left( t \delta \right)  &\!\! 0\\
\\[-8pt]
0 &\!\! \cos \! \left( t \bar{\delta} \right) 
\EA \!\!\!
\right]
\!\!+\!\!
\left[ \!\!\!
\BA{cc}
\delta^{-1} &\!\! 0\\
\\[-8pt]
0 &\!\! \bar{\delta}^{-1}
\EA \!\!\!
\right] \!\!
\left[ \!\!\!
\BA{cc}
\sin \! \left( t \delta \right)  &\!\! 0\\
\\[-8pt]
0 &\!\! \sin \! \left( t \bar{\delta} \right) 
\EA \!\!\!
\right]
\!\cdot\!
{\displaystyle \frac{{\cal F}}{i\hbar }} 
\left( ||\delta^2||,  |||\bar{\delta}^2|| \!>\! 0 \right) .
\label{expRiccatiMat3}
\ea
\vspace{-0.1cm}
\ba
\exp\{t{\displaystyle \frac{{\cal F}}{i\hbar }}\}
\!\!=\!\!
\left[ \!\!\!
\BA{cc}
\cosh \! \left( t \delta \right)  &\!\! 0\\
\\[-8pt]
0 &\!\! \cosh \! \left( t \bar{\delta} \right) 
\EA \!\!\!
\right]
\!\!+\!\!
\left[ \!\!\!
\BA{cc}
\delta^{-1} &\!\! 0\\
\\[-8pt]
0 &\!\! \bar{\delta}^{-1}
\EA \!\!\!
\right] \!\!
\left[ \!\!\!
\BA{cc}
\sinh \! \left( t \delta \right)  &\!\! 0\\
\\[-8pt]
0 &\!\! \sinh \! \left( t \bar{\delta} \right) 
\EA \!\!\!
\right]
\!\cdot\!
{\displaystyle \frac{{\cal F}}{i\hbar }} 
\left(-||\delta^2||, -||\bar{\delta}^2|| \!<\! 0 \right) .
\label{expRiccatiMat4}
\ea\\[-6pt]
In the above
we suppose the existence o the inverse matrices
$\delta^{-1}$ and $\bar{\delta}^{-1}$.
Following also the Inoguchi method
\cite{Inoguchi.01},
the solutions
(\ref{expRiccatiMat3})
and
(\ref{expRiccatiMat4})
are shown to satisfy the Riccati equation
as follows:\\[-22pt]
\ba
\!\!\!\!
\BA{ll}
\left. {\displaystyle \frac{d}{dt }} \! \right|_{t \!=\! 0} \!\!
q \! \left( \! T_{\exp\{ \! tA(t)\} \! } \! (\!q\!) \! \right)
\!\!=\!\!
\left. {\displaystyle \frac{d}{dt }} \! \right|_{t \!=\! 0} \!\!
{\displaystyle
\frac{
\BA{c}
\cos \! \left( \! t \delta \! \right)
\!+\!
\delta^{-1} \!
\sin \! \left( \! t \delta \! \right) \!
{\displaystyle \frac{F}{i\hbar }}
\cdot
q
~~+\! \\
\\[-12pt]
\cosh \! \left( \! t \delta \! \right) 
\!\!+\!\!
\delta^{-1} \!
\sinh \! \left( \! t \delta \! \right) \!
{\displaystyle \frac{F}{i\hbar }} 
\!\cdot\!
q
~+\!
\EA \!\!
\BA{c}
\delta^{-1} \!
\sin \! \left( \! t \delta \! \right) \!
{\displaystyle \frac{D}{i\hbar }} \\
\\[-10pt]
\delta^{-1} \!
\sinh \! \left( \! t \delta \! \right) \!
{\displaystyle \frac{D}{i\hbar }} 
\EA \!\!\! 
}
{
\BA{c}
\cos \! \left( \! t \bar{\delta} \! \right)
\!-\!
\bar{\delta}^{-1} \!
\sin \! \left( \! t \bar{\delta} \! \right) \!
{\displaystyle \frac{\bar{F}}{i\hbar }}  \\
\\[-12pt]
\cosh \! \left( \! t \bar{\delta} \! \right)
\!-\!
\bar{\delta}^{-1} \!
\sinh \! \left( \! t \bar{\delta} \! \right) \!
{\displaystyle \frac{\bar{F}}{i\hbar }} 
\EA \!\!\!
\BA{c}
\!\!\!\!\!\!
-~
\bar{\delta}^{-1} \!
\sin \! \left( \! t \bar{\delta} \! \right) \!
{\displaystyle \frac{\bar{D}}{i\hbar }}
\!\cdot\!
q   \\
\\[-10pt]
\!-~
\bar{\delta}^{-1} \!
\sinh \! \left( \! t \bar{\delta} \! \right) \!
{\displaystyle \frac{\bar{D}}{i\hbar }}
\!\cdot\!
q 
\EA \!\!\!
}
}
\!=\!
{\displaystyle
\frac{D}{i\hbar } 
\!+\!
\frac{F}{i\hbar } q
\!+\!
q\frac{\bar{F}}{i\hbar }
\!+\!
q \frac{\bar{D}}{i\hbar } q
} ,
\EA
\label{RiccatiSol2}
\ea\\[-10pt]
from which,
we can surely prove that they satisfy the Riccati equation for $q(t)$
(\ref{RiccatiEq}).
The case $||\delta^2|| \!=\! 0$
(\ref{expRiccatiMat2})
is also easily proved.
The relations
$
FD
\! =\!
D\bar{F}
$
and
$
\bar{F}\bar{D}
\! =\!
\bar{D}F
$
play a crucial role to derive
(\ref{expRiccatiMat3})
and
(\ref{expRiccatiMat4}).
$\!\!$They are also satisfied trivially
for the previous simple case
(\ref{RiccatiMat}).

The above construction of the solution is deeply connected to
the way of construction developed by
Berceanu, Gheorghe and de Monvel,
who have asserted that
the matrix Riccati equation is a {\it flow} on the Grassmann coset manifold
\cite{BG.91}.\\[-14pt]

We use a $N \times 2N$ isometric matrix
$
u
\!=\!\!
\left[ \!\!
\BA{c}
b \\
\\[-14pt]
a 
\EA \!\!
\right] 
$
and a matrix $g$ given by
$
g
\!=\!\! 
\left[ \!\!
\BA{cc}
a &\!\!\bar{b} \\
\\[-6pt]
b &\!\! \bar{a} \!\!\!
\EA 
\right] \!
(\det \! g \! =\! 1) 
$,\\[-4pt]
given by the first of Eq.
(\ref{calGprime}).
Let us suppose the following HB eigenvalue equations:\\[-16pt]
\ba
\!\!
{\cal F} \!
\left[ \!\!
\BA{c}
b \\
\\[-14pt]
a 
\EA \!\!
\right]_{\!i}
\!\!=\!\!
- \!
\left[ \!\!
\BA{c}
b \\
\\[-14pt]
a 
\EA \!\!
\right]_{\!i} \!
\epsilon_i ,
\mbox{or}~
g^{\mbox{\scriptsize T}} \!\!
{\cal F}
\!\!=\!\!
\left[ \!\!
\BA{cc}
\epsilon & 0 \!\!\\
\\[-6pt]
0 &\!\! -\epsilon \!\!
\EA 
\right] \!
g^{\mbox{\scriptsize T}},
\epsilon
\!\equiv\!
\mbox{diagonal matrix}(\epsilon_1, \cdots, \epsilon_N) ,
\det \!\! {\cal F}
\!\!=\!\! 
(\!-1)^N \!\! \left( \det \! \epsilon \right)^2 \! .
\label{HBeigeneq}
\ea\\[-12pt]
Keeping the form of equation
(\ref{HBeigeneq}),
we derive a usual TDHB equation
in the form as\\[-16pt]
\ba
i\hbar \!
\left[ \!\!
\BA{c}
\dot{\bar{b}} \\
\\[-14pt]
\dot{\bar{a}} 
\EA \!\!
\right]
\!=\!
- \bar{{\cal F}} \!
\left[ \!\!
\BA{c}
\bar{b} \\
\\[-10pt]
\bar{a} 
\EA \!\!
\right] ,
~\mbox{or}~
i\hbar \!
\left[ \!\!
\BA{c}
\dot{a} \\
\\[-12pt]
\dot{b} 
\EA \!\!
\right]
\!=\!
- \bar{{\cal F}} \!
\left[ \!\!
\BA{c}
a \\
\\[-10pt]
b 
\EA \!\!
\right] ,~
\left[ \!\!
\BA{cc}
0 & 1_N \!\!\\
\\[-6pt]
1_N &\!\! 0 \!\!
\EA 
\right] \!\!
\left( -\bar{{\cal F}} \right) \!\!
\left[ \!\!
\BA{cc}
0 & 1_N \!\!\\
\\[-6pt]
1_N &\!\! 0 \!\!
\EA 
\right] \!
\!=\!
{\cal F},
\label{HBeigeneq3}
\ea\\[-10pt]
which are written in more compact forms as\\[-18pt]
\ba
i\hbar 
\dot{\bar{g}}
\!=\!
{\cal F}
\bar{g},
~\mbox{or}~
i\hbar 
\dot{g}^{\mbox{\scriptsize T}} \!
\!=\!
-
g^{\mbox{\scriptsize T}} \!
{\cal F} .
\label{HBeigeneq4}
\ea


\newpage

\setcounter{equation}{0}
\renewcommand{\theequation}{\arabic{section}.\arabic{equation}}

\section{Concluding remarks and further outlook}

~~~~In this paper
first
we present the induced representation of
$SO(2N)$ canonical transformation group
and introduce $\frac{SO(2N)}{U(N)}$ coset variables
\cite{Nishi.81}.
We give the derivation of the TDHB equation from the
Euler-Lagrange equation of motion for the variables in the TDSCF.
The TDHB theory is a powerful tool for superconducting fermion systems
\cite{Bog.59,RS.80}. 
The TDHB WF represents
the $\!$TD behavior $\!$of Bose condensate of fermion pairs.
$\!\!$It is a good approximation for
the ground state $\!$of a fermion system with a pairing interaction,
producing the spontaneous Bose condensation.
$\!\!$The TDHB equation on the K\"{a}hler coset space
$\!\frac{G}{H} \!\!=\!\! \frac{SO(2N)}{U(N)} \!$ is derived
from the Euler-Lagrange
equation of motion for the coset variables.
To describe the classical on the coset manifold,
we start from the local equation of motion
\cite{Nishi.81,BG.91}.
This equation becomes of the Riccati type
\cite{Riccati.1758,Reid.72,Z_Itkin.73}.
After providing a simple two-level model and a solution for a coset variable,
we can give successfully a general solution of the TDRHB equation
for the coset variables.

Next,
along the same strategy developed by van Holten et al.
\cite{NNH.01},
we have extended to the SUSY $\sigma$-model
on the K\"{a}hler coset space
$\!\frac{G}{H} \!=\!\! \frac{SO(2N \!+\! 2)}{U(N \!+\! 1)}\!$
based on the $SO(2N \!\!+\!\! 1)$ Lie algebra of fermion operators
\cite{SJCF.08}.
Following Fukutome,
by embedding a $\!SO(2N \!\!+\!\! 1)\!$ group into an anomaly-free spinor rep of
$\!SO(2\!N \!+\! 2)\!$ group and
using $\!\frac{SO(2N \!+\! 2)}{U(N \!+\! 1)}\!$ coset variables
\cite{Fuk.77,Fuk.81},
we have studied a new aspect of the extended anomaly-free SUSY $\sigma$-model
and have given a corresponding K\"{a}hler potential and
then a Killing potential
based on a positive chiral spinor rep.
The theory is invariant under a SUSY transformation
and the Killing potential is expressed in terms of the coset variables.
Using such mathematical manipulation,
we have constructed
the generalized density matrix in the $SO(2N)$ CS rep
and obtained the Harish-Chandra decomposition
for the $SO(2N)$ matrix
based on the nonlinear M\"{o}bius transformation together with
the geodesic flow on the manifold.
Further using
an anomaly-free spinor rep of the $SO(2N)$ group,
we have obtained
an irreducible completely antisymmetric $SU(N)$-tensor
and -dual tensor under the transformation
and then have derived the corresponding Killing potential
for such the tensors.

The generalized density matrix $W~( \!=\! uu^\dag )$
(\ref{densitymat})
is expressed in term of the matrix $g$
as\\[-16pt]
\ba
W
\!=\!
{\displaystyle \frac{1}{2}}
\bar{g} \!\!
\left[ \!\!
\BA{cc}
-1_N & 0 \!\!\\
\\[-6pt]
0 &\!\! 1_N \!\!
\EA 
\right] \!\!
g^{\mbox{\scriptsize T}}
\!+\!
{\displaystyle \frac{1}{2}} \!
\left[ \!\!
\BA{cc}
-1_N & 0 \!\!\\
\\[-6pt]
0 &\!\! 1_N \!\!
\EA 
\right] ,~
W^2
\!=\!
W .
\label{densitymat2}
\ea\\[-12pt]
Using
(\ref{HBeigeneq4}),
the equation of motion for $W$
is given as\\[-18pt]
\ba
\BA{ll}
i\hbar 
\dot{W}
\!\!\!\!&\!=\!
i\hbar
{\displaystyle \frac{1}{2}}
\dot{\bar{g}} \!\!
\left[ \!\!
\BA{cc}
-1_{\!N} & 0 \!\!\\
\\[-6pt]
0 &\!\! 1_{\!N} \!\!
\EA 
\right] \!\!
g^{\mbox{\scriptsize T}}
\!+\!
i\hbar
{\displaystyle \frac{1}{2}}
\bar{g} \!\!
\left[ \!\!
\BA{cc}
-1_{\!N} & 0 \!\!\\
\\[-6pt]
0 &\!\! 1_{\!N} \!\!
\EA 
\right] \!\!
\dot{g}^{\mbox{\scriptsize T}}
\!=\!
{\cal F}
{\displaystyle \frac{1}{2}}
\bar{g} \!\!
\left[ \!\!
\BA{cc}
-1_{\!N} & 0 \!\!\\
\\[-6pt]
0 &\!\! 1_{\!N} \!\!
\EA 
\right] \!\!
g^{\mbox{\scriptsize T}}
\!-\!
{\cal F}
{\displaystyle \frac{1}{2}}
\bar{g} \!\!
\left[ \!\!
\BA{cc}
-1_N & 0 \!\!\\
\\[-6pt]
0 &\!\! 1_N \!\!
\EA 
\right] \!\!
g^{\mbox{\scriptsize T}} \\
\\[-10pt]
&\!=\!
{\cal F} \!\!
\left( \!
W
\!-\!
{\displaystyle \frac{1}{2}} \!
\!\cdot\!
1_{\!2N} \!\!
\right) \!
\!-\!
\left( \!
W
\!-\!
{\displaystyle \frac{1}{2}} \!
\!\cdot\!
1_{\!2N} \!\!
\right) \!\!
{\cal F}
\!=\!
[ {\cal F}, W ] .
\EA
\label{eqmotion_densitymat}
\ea\\[-10pt]
This is just another famous form of the TDHB equation.
According to Berceanu et.al.
\cite{BG.91},
the nonlinear M\"{o}bius transformation
(\ref{MtransmatQ})
given by
$\exp\{t \frac{{\cal F}}{i\hbar }\}$
(\ref{expRiccatiMat2}), (\ref{expRiccatiMat3})
and
(\ref{expRiccatiMat4})
act on the $\frac{SO(2N)}{U(N)}$ coset variables $q$
and then produces the solution of the TDRHB equation.
How does the
$\exp\{t \frac{{\cal F}}{i\hbar }\}$
act on the generalized density matrix $W$?
This may be made as follows:
\beq
{}^{\exp\{ t\frac{{\cal F}}{i\hbar }\}}\! W
\!\!=\!\!
\left[ \!\!\!
\BA{cc} 
{}^{\exp\{ t\frac{{\cal F}}{i\hbar }\}}\! R & {}^{\exp\{ t\frac{{\cal F}}{i\hbar }\}}\! K \\
\\[-6pt]
-{}^{\exp\{ t\frac{{\cal F}}{i\hbar }\}}\! \bar{K} &
{}^{\exp\{ t\frac{{\cal F}}{i\hbar }\}} \! ( 1_{\!N} \!-\! \bar{R})
\EA \!\!\!
\right] \! ,
\BA{c}
{}^{\exp\{ t\frac{{\cal F}}{i\hbar }\}}\! R
\!=\!
{}^{\exp\{ t\frac{{\cal F}}{i\hbar }\}} \! (1_{\!N} \!-\! \chi) ,\\
\\[-6pt]
{}^{\exp\{ t\frac{{\cal F}}{i\hbar }\}}\! K
\!=\!
{}^{\exp\{ t\frac{{\cal F}}{i\hbar }\}} \! (\chi q)  ,
\EA
\chi
\!=\!
( 1_{\!N} \!+\! qq^\dag )^{-1} ,
\label{Fact_densitymat}
\eeq
where we have used the relations
(\ref{matRandK}).
The calculation of
$
\left. {\displaystyle \frac{d}{dt }} \! \right|_{t \!=\! 0} \!\!
W \!\! \left( \! T_{\exp\{ t \frac{{\cal F}}{i\hbar }\}} ( W ) \! \right)
$
seems to be a very hard task.
What is the meaning of this result?
Does the result open a new field in theoretical physics?
This is an interesting future problem to be solved.


\newpage

\leftline{\large{\bf Appendix}}
\appendix


\def\thesection{\Alph{section}}
\setcounter{equation}{0}
\renewcommand{\theequation}{\Alph{section}.\arabic{equation}}
\section{Bosonization of SO(2N) Lie operators}


~~~Consider 
a fermion state vector $\ket \Psi$ 
corresponding to a function 
$\Psi (g)$ in $g \in SO(2N)$:  
\beqa
\BA{l}
\ket \Psi
=
\int \! U (g) \ket 0 \bra 0 U^\dag (g) \ket \Psi dg
=
\int \! U (g) \ket 0 \Psi (g) dg .
\EA
\label{statePsi}
\eeqa
The $g$ is given by
(\ref{calGprime})
and
the $dg$ is an invariant group integration.  
When an infinitesimal operator 
$\mathbb{I}_g \!+\! \delta \widehat{g}$ and
a corresponding infinitesimal unitary operator 
$U (1_{\!2N} \!+\! \delta g)$
is operated on $\ket \Psi$,
using
$U^{-1}(1_{\!2N} \!+\! \delta g) 
\!=\! 
U (1_{\!2N} \!-\! \delta g)$, 
it transforms $\ket \Psi$ as
\beqa
\!\!\!\!\!\!
\BA{ll}
&U (1_{\!2N} - \delta g)  \ket \Psi
=
(\mathbb{I}_g - \delta \widehat{g}) \ket \Psi
=
\int \! U (g) \ket 0 \bra 0 
U^\dag ((1_{\!2N} + \delta g)g) \ket \Psi dg \\
\\[-8pt]
&=
\int \! U (g) \ket 0 \Psi 
((1_{\!2N} 
+ 
\delta g){\cal G}) 
d{\cal G} 
=
\int \! U (g) \ket 0 (1_{\!2N} + \delta \mbox{\boldmath $g$})  
\Psi (g) dg,
\EA
\label{infinitrans}
\eeqa 
\beqa
\!\!\!\!\!\!
\left.
\BA{ll}
&
1_{\!2N} + \delta g
=\!
\left[ \!\!
\BA{cc} 
1_{\!N} + \delta a & \delta \bar{b}\\
\\[-4pt]
\delta b & 1_{\!N} + \delta \bar{a} \\ 
\EA \!\!
\right] ,~
\delta a^\dag = - \delta a,~
\mbox{tr}\delta a = 0 ,~
\delta b = - \delta b^{\mbox{\scriptsize T}} , \\
\\[-8pt]
&\delta \widehat{g}
=
\delta a^\alpha_{~\beta} E^\beta_{~\alpha}
+
{\displaystyle \frac{1}{2}} 
\left(
\delta b_{\alpha\beta} E^{\beta\alpha}
+
\delta \bar{b}_{\alpha\beta} E_{\beta\alpha}
\right) ,~
\delta \mbox{\boldmath $g$}
=
\delta a^\alpha_{~\beta} \mbox{\boldmath $E^\beta_{~\alpha}$}
+
{\displaystyle \frac{1}{2}} 
\left(
\delta b_{\alpha\beta} \mbox{\boldmath $E^{\beta\alpha}$}
+ 
\delta \bar{b}_{\alpha\beta} \mbox{\boldmath $E_{\beta\alpha}$}
\right) . \!\!
\EA
\right\}
\label{infinitesimalop}
\eeqa
Equation
(\ref{infinitrans})
shows that
the operation of $\mathbb{I}_g \!-\! \delta \widehat{g}$ 
on the $\ket \Psi$ in the fermion space 
corresponds to the left multiplication by $1_{\!2N} \!+\! \delta g$
for the variable of the $g$ of the function $\Psi (g)$.
For a small parameter $\epsilon$,
we obtain a representation on the $\Psi (g)$ as
\beq
\rho (e^{\epsilon \delta g}) \Psi (g)
=
\Psi (e^{\epsilon \delta g} g)
=
\Psi (g + \epsilon \delta g g)
=
\Psi (g + dg) ,
\label{reponPsi}
\eeq
which leads us to a relation 
$dg 
= 
\epsilon \delta g $.
From this,
we express it explicitly as,
\beqa
\left.
\BA{ll}
&
dg
=\!
\left[ \!\!
\BA{cc} 
da & d\bar{b} \\
\\[-4pt]
db & d\bar{a} \\ 
\EA \!\!
\right]
=
\epsilon \!
\left[ \!\!
\BA{cc} 
\delta a a + \delta \bar{b} b  & 
\delta a \bar{b} + \delta \bar{b} \bar{a} \\
\\[-4pt]
\delta b a + \delta \bar{a} b & 
\delta \bar{a} \bar{a} + \delta b \bar{b} \\ 
\EA \!\!
\right] ,\\
\\
&
da
=
\epsilon
{\displaystyle \frac{\partial a}{\partial \epsilon }}
=
\epsilon
(\delta a a + \delta \bar{b} b) ,~~
db
=
\epsilon
{\displaystyle \frac{\partial b}{\partial \epsilon }}
=
\epsilon
(\delta b a + \delta \bar{a} b) .
\EA
\right\}
\label{dGdAdB} 
\eeqa
A differential representation of $\rho (\delta g)$,
$d\rho (\delta g)$,
is given as
\beq
d\rho (\delta g) \Psi (g)
=\!
\left[
\frac{\partial a^\alpha_{~\beta}}{\partial \epsilon }
\frac{\partial }{\partial a^\alpha_{~\beta} }
+
\frac{\partial b_{\alpha\beta}}{\partial \epsilon }
\frac{\partial }{\partial b_{\alpha\beta} }
+
\frac{\partial \bar{a}^\alpha_{~\beta}}{\partial \epsilon }
\frac{\partial }{\partial \bar{a}^\alpha_{~\beta} }
+
\frac{\partial \bar{b}_{\alpha\beta}}{\partial \epsilon }
\frac{\partial }{\partial \bar{b}_{\alpha\beta}}
\right] \!
\Psi (g) .
\label{diffrep} 
\eeq
Substituting
(\ref{dGdAdB})
into
(\ref{diffrep}),
we can get explicit forms of the differential representation
\beq
d\rho (\delta g) \Psi (g)
=
\delta \mbox{\boldmath $g$}
\Psi (g),
\eeq
where each operator in $\delta \mbox{\boldmath $g$}$
is expressed in a differential form as
\beqa
\left.
\BA{ll}
&
\mbox{\boldmath $E^\alpha_{~\beta}$}
=
\bar{b}_{\alpha\gamma}
{\displaystyle \frac{\partial }{\partial \bar{b}_{\beta\gamma}}}
-
b_{\beta\gamma}
{\displaystyle \frac{\partial }{\partial b_{\alpha\gamma}}}
-
\bar{a}^{\beta}_{~\gamma}
{\displaystyle \frac{\partial }{\partial \bar{a}^{\alpha}_{~\gamma}}}
+
a^\alpha_{~\gamma}
{\displaystyle \frac{\partial }{\partial a^\beta_{~\gamma}}} 
=
\mbox{\boldmath $E^{\beta \dag }_{~\alpha}$} ,\\
\\[-10pt]
&
\mbox{\boldmath $E_{\alpha\beta}$}
=
\bar{a}^{\alpha}_{~\gamma}
{\displaystyle \frac{\partial }{\partial \bar{b}_{\beta\gamma}}}
-
b_{\beta\gamma}
{\displaystyle \frac{\partial }{\partial a^\alpha_{~\gamma}}}
-
\bar{a}^{\beta}_{~\gamma}
{\displaystyle \frac{\partial }{\partial \bar{b}_{\alpha\gamma}}}
+
b_{\alpha\gamma}
{\displaystyle \frac{\partial }{\partial a^\beta_{~\gamma}}} 
=
\mbox{\boldmath $E^{\beta\alpha \dag }$} ,\\
\\[-8pt]
&
\mbox{\boldmath $E^{\alpha \dag }_{~\beta}$}
=
-
\mbox{\boldmath $\bar{E}^{\alpha}_{~\beta}$},~~~~~~~
\mbox{\boldmath $E_{\alpha\beta }^\dag $}
=
-
\mbox{\boldmath $\bar{E}_{\alpha\beta }$},~~~~~~~
\mbox{\boldmath $E_{\alpha\beta}$}
=
-
\mbox{\boldmath $E_{\beta\alpha}$} .
\EA
\right\}
\label{diffops}
\eeqa
We define the boson operators 
$\mbox{\boldmath $a^\alpha_{~\beta}$}$,
$\mbox{\boldmath $\bar{a}^{\alpha}_{~\beta}$}$, etc.,
from every variable
$a^\alpha_{~\beta}$,
$\bar{a}^{\alpha}_{~\beta}$, etc.,
as\\[-12pt]
\beqa
\left.
\BA{ll}
&
\mbox{\boldmath $a$}
\stackrel{\mathrm{def}}{=}
{\displaystyle
\frac{1}{\sqrt{2}} \!
\left( \!
a \!+\! \frac{\partial }{\partial \bar{a} } \!
\right) 
} ,~
\mbox{\boldmath $a^\dag$}
\stackrel{\mathrm{def}}{=}
{\displaystyle
\frac{1}{\sqrt{2}} \!
\left( \!
\bar{a} \!-\! \frac{\partial }{\partial a} \!
\right) 
} ,~
\mbox{\boldmath $\bar{a}$}
\stackrel{\mathrm{def}}{=}
{\displaystyle
\frac{1}{\sqrt{2}} \!
\left( \!
\bar{a} \!+\! \frac{\partial }{\partial a} \!
\right) 
} ,~
\mbox{\boldmath $a^{\mbox{\scriptsize T}}$}
\stackrel{\mathrm{def}}{=}
{\displaystyle
\frac{1}{\sqrt{2}} \!
\left( \!
a \!-\! \frac{\partial }{\partial \bar{a} } \!
\right) 
} , \\
\\[-4pt]
&
[\mbox{\boldmath $a$},~\mbox{\boldmath $a^\dag$}]
=
1 ,~~
[\mbox{\boldmath $\bar{a}$},~
\mbox{\boldmath $a^{\mbox{\scriptsize T}}$}]
=1 ,~~
[\mbox{\boldmath $a$},~\mbox{\boldmath $\bar{a}$}]
=
[\mbox{\boldmath $a$},~
\mbox{\boldmath $a^{\mbox{\scriptsize T}}$}]
=
0 ,~~
[\mbox{\boldmath $a$}^\dag ,~\mbox{\boldmath $\bar{a}$}]
=
[\mbox{\boldmath $a$}^\dag ,~
\mbox{\boldmath $a^{\mbox{\scriptsize T}}$}]
=
0 ,
\EA
\right\}
\label{boseops}
\eeqa
where $a$ is a complex variable.
Similar definitions hold for $b$ in order
to define the boson operators 
$\mbox{\boldmath $b_{\alpha\beta}$}$,
$\mbox{\boldmath $\bar{b}_{\alpha\beta}$}$, etc.
By noting the relations
$~
{\displaystyle
\bar{a} \frac{\partial }{\partial \bar{a} }
\!-\!
a \frac{\partial }{\partial a}
\!=\!
\mbox{\boldmath $a^\dag$}
\mbox{\boldmath $a$}
\!-\!
\mbox{\boldmath $a^{\mbox{\scriptsize T}}$}
\mbox{\boldmath $\bar{a}$} 
}
$
and
$
{\displaystyle
\bar{a} \frac{\partial }{\partial \bar{b}
}
\!-\!
b \frac{\partial }{\partial a}
\!=\!
\mbox{\boldmath $a^\dag$}
\mbox{\boldmath $b$}
\!-\!
\mbox{\boldmath $b^{\mbox{\scriptsize T}}$}
\mbox{\boldmath $\bar{a}$}},
$
the differential operators
(\ref{diffops})
can be converted into a boson operator representation
\beqa
\left.
\BA{ll}
&
\mbox{\boldmath $E^\alpha_{~\beta}$}
=
\mbox{\boldmath $b^\dag _{\alpha\gamma}$}
\mbox{\boldmath $b_{\beta\gamma}$}
-
\mbox{\boldmath $b^{\mbox{\scriptsize T}}_{\beta\gamma}$}
\mbox{\boldmath $\bar{b}_{\alpha\gamma}$}
-
\mbox{\boldmath $a^{\beta \dag }_{~\gamma}$}
\mbox{\boldmath $a^\alpha_{~\gamma}$}
+
\mbox{\boldmath $a^{p \mbox{\scriptsize T}}_{~r}$}
\mbox{\boldmath $\bar{a}^{q}_{~r}$} 
=
\mbox{\boldmath $b^\dag _{p \tilde{r}}$}
\mbox{\boldmath $b_{q \tilde{r}}$}
-
\mbox{\boldmath $a^{\beta \dag }_{~\tilde{\gamma}}$}
\mbox{\boldmath $a^\alpha_{~\tilde{\gamma}}$} ,\\
\\[-6pt]
&
\mbox{\boldmath $E_{\alpha\beta}$}
=
\mbox{\boldmath $a^{\alpha \dag }_{~\gamma}$}
\mbox{\boldmath $b_{\beta\gamma}$}
-
\mbox{\boldmath $b^{\mbox{\scriptsize T}}_{\beta\gamma}$}
\mbox{\boldmath $\bar{a}^{\alpha}_{~\gamma}$}
-
\mbox{\boldmath $a^{q \dag }_{~\gamma}$}
\mbox{\boldmath $b_{\alpha\gamma}$}
+
\mbox{\boldmath $b^{\mbox{\scriptsize T}}_{\alpha\gamma}$}
\mbox{\boldmath $\bar{a}^{\beta}_{~\gamma}$} 
=
\mbox{\boldmath $a^{\alpha \dag }_{~\tilde{\gamma}}$}
\mbox{\boldmath $b_{\beta \tilde{\gamma}}$}
-
\mbox{\boldmath $a^{\beta \dag }_{~\tilde{\gamma}}$}
\mbox{\boldmath $b_{\alpha \tilde{\gamma}}$} ,
\EA
\right\}
\label{bosonizesops}
\eeqa
by using the notation
$\!
\mbox{\boldmath $a^{\alpha \mbox{\scriptsize T}}_{~\gamma\!+\!N}$} 
\!\!=\!\!
\mbox{\boldmath $b^\dag _{\alpha\gamma}$}\!
$
and
$\!
\mbox{\boldmath $b^{\mbox{\scriptsize T}}_{\alpha\gamma\!+\!N}$}
\!\!=\!\!
\mbox{\boldmath $a^{\alpha \dag }_{~\gamma}$}\!
$
to use a suffix $\tilde{\gamma}$ running from 
1 to $N$ and from $N$ to $2N$.
Then we have the boson images of 
the fermion $SO(2N)$ Lie operators as\\[-12pt]
\beqa
\BA{c}
\mbox{\boldmath $E^\alpha_{~\beta }$}
=
\mbox{\boldmath $b^\dag _{\alpha \tilde{\gamma}}$}
\mbox{\boldmath $b_{\beta \tilde{\gamma}}$}
-
\mbox{\boldmath $a^{\beta \dag }_{~\tilde{\gamma}}$}
\mbox{\boldmath $a^\alpha_{~\tilde{\gamma}}$} ,~~
\mbox{\boldmath $E_{\alpha \beta }$}
=
\mbox{\boldmath $a^{\alpha \dag }_{~\tilde{\gamma}}$}
\mbox{\boldmath $b_{\beta \tilde{\gamma}}$}
-
\mbox{\boldmath $a^{\beta \dag }_{~\tilde{\gamma}}$}
\mbox{\boldmath $b_{\alpha \tilde{\gamma}}$} .
\EA
\label{bosonimage2}
\eeqa\\[-26pt]

Using the relations
$~
{\displaystyle \frac{\partial }{\partial a^\alpha_{~\beta}}}
\det a
\!=\!
(a^{-1})^{~\beta}_\alpha
\det a
$
and
$
{\displaystyle \frac{\partial }{\partial a^\alpha_{~\beta}}}
(a^{-1})^{~\gamma}_\delta
\!=\!
-
(a^{-1})^{~\beta}_\delta
(a^{-1})^{~\gamma}_\alpha 
$,
we get the relations which are valid when operated on functions
on the right coset$\frac{SO(2N)}{SU(N)}$\\[-12pt]
\beqa
\left.
\BA{ll}
&
{\displaystyle \frac{\partial }{\partial b_{\alpha\beta}}}
=
\sum _{\gamma < \alpha}
(a^{-1})^{~\beta}_\gamma
{\displaystyle \frac{\partial }{\partial q_{\alpha\gamma}}},\\
\\[-12pt]
&
{\displaystyle \frac{\partial }{\partial a^\alpha_{~\beta}}}
=
-
\sum _{\delta < \gamma < \alpha}
q_{\gamma\alpha}
(a^{-1})^{~\beta}_\delta 
{\displaystyle \frac{\partial }{\partial q_{\gamma\delta}}}
-
{\displaystyle \frac{i}{2}}
(a^{-1})^{~\beta}_\alpha 
{\displaystyle \frac{\partial }{\partial \tau }} ,
\EA
\right\}
\label{differentialformulas}
\eeqa
using which,
the bosonized operators
(\ref{bosonizesops}) 
are 
expressed by the closed first order differential form
over the $\!\frac{SO(2N)}{U(N)}\!$ coset space
in terms of 
the coset variables $q_{\alpha\beta}$ 
and 
a phase variable $\tau$.



\def\thesection{\Alph{section}}
\setcounter{equation}{0}
\renewcommand{\theequation}{\Alph{section}.\arabic{equation}}
\section{Vacuum function for bosons}


~
We show that function 
$\Phi_{00}(g) (\!\!=\!\! \overline{\bra \! 0 U(g)  \! \ket 0})$ in $g \!\!\in\!\! SO(2N)$
corresponds to the free fermion vacuum function 
in the physical fermion space.
Then the
$\Phi_{00}(g)$
must satisfy the conditions\\[-8pt]
\beq
\left( \!\!
\mbox{\boldmath $E^\alpha_{~\beta}$}
+
{\displaystyle \frac{1}{2}\delta_{\alpha\beta}} \!\!
\right) \!
\Phi_{00}(g)
=
0 ,~~
\Phi_{00}(1_{\!2N})
=
1 .
\label{vacuumcondition}
\eeq
The vacuum function $\Phi_{00}(g)$ which satisfy
(\ref{vacuumcondition})
is given by
$
\Phi_{00}(g) 
\!=\!
[\det (\bar{a})]^{\frac{1}{2}} ,
$
the proof of which is made easily as follows:\\[-12pt]
\beqa
\!\!\!\!\!\!\!\!
\BA{ll}
&
\left( \!\!
\mbox{\boldmath $E^\alpha_{~\beta}$}
+
{\displaystyle \frac{1}{2}\delta_{\alpha\beta}} \!\!
\right) \!
[\det (\bar{a})]^{\frac{1}{2}}
\!=\!
{\displaystyle \frac{1}{2}\delta_{\alpha\beta}}
[\det (\bar{a})]^{\frac{1}{2}}
\!\!+\!\!
\left( \!\!
\bar{b}_{\alpha\gamma}
{\displaystyle \frac{\partial }{\partial \bar{b}_{\beta\gamma}}}
\!-\!
b_{\beta\gamma}
{\displaystyle \frac{\partial }{\partial b_{\alpha\gamma}}}
\!-\!
\bar{a}^{\beta}_{~\gamma}
{\displaystyle \frac{\partial }{\partial \bar{a}^{\alpha}_{~\gamma}}}
\!+\!
a^\alpha_{~\gamma}
{\displaystyle \frac{\partial }{\partial a^\beta_{~\gamma}}} \!\!
\right) \!
[\det (\bar{a})]^{\frac{1}{2}} \\
\\[-12pt]
&
=
{\displaystyle \frac{1}{2}\delta_{\alpha\beta}}
[\det (\bar{a})]^{\frac{1}{2}}
\!-\!
\bar{a}^{\beta}_{~\gamma}
{\displaystyle \frac{\partial }{\partial \bar{a}^{\alpha}_{~\gamma}}}
[\det (\bar{a})]^{\frac{1}{2}} 
\!=\!
{\displaystyle \frac{1}{2}\delta_{\alpha\beta}}
[\det (\bar{a})]^{\frac{1}{2}}
\!-\!
{\displaystyle 
\frac{1}{2}\frac{1}{[\det (\bar{a})]^{\frac{1}{2}}}}\!
\bar{a}^{\beta}_{~\gamma}\!
{\displaystyle \frac{\partial }{\partial \bar{a}^{\alpha}_{~\gamma}}}\!
\det (\bar{a}) \\
\\[-12pt]
&
=
{\displaystyle \frac{1}{2}\delta_{\alpha\beta}}
[\det (\bar{a})]^{\frac{1}{2}}
-
{\displaystyle 
\frac{1}{2}\frac{1}{[\det (\bar{a})]^{\frac{1}{2}}}}
(\bar{a}\bar{a}^{-1})_{\beta\alpha}
\det (\bar{a}) 
= 0 ,
\EA
\label{vacuum1}
\eeqa
\vspace{-0.3cm}
\beqa
\!\!\!\!\!\!\!\!\!\!\!\!
\mbox{\boldmath $E_{\alpha\beta}$}
[\det (\bar{a})]^{\frac{1}{2}} 
\!=\!
\left( \!
\bar{a}^{\alpha}_{~\gamma}
{\displaystyle \frac{\partial }{\partial \bar{b}_{\beta\gamma}}}
-
b_{\beta\gamma}
{\displaystyle \frac{\partial }{\partial a^\alpha_{~\gamma}}}
-
\bar{a}^{\beta}_{~\gamma}
{\displaystyle \frac{\partial }{\partial \bar{b}_{\alpha\gamma}}}
+
b_{\alpha\gamma}
{\displaystyle \frac{\partial }{\partial a^\beta_{~\gamma}}} \!
\right) \!
[\det (\bar{a})]^{\frac{1}{2}} 
= 0 .
\label{vacuum2}
\eeqa
Thus the vacuum function $\Phi_{00}(g)$ in $g \!\in\! SO(2N)$ satisfies
$
\left( \!\!
\mbox{\boldmath $E^\alpha_{~\beta }$}
\!+\!
\frac{1}{2}\delta_{\alpha \beta } \!\!
\right) \!
\Phi_{00}(g)
\!=\!
\mbox{\boldmath $E_{\alpha \beta }$}\Phi_{00}(g)
\!=\!
0 
$.

\newpage


\def\thesection{\Alph{section}}
\setcounter{equation}{0}
\renewcommand{\theequation}{\Alph{section}.\arabic{equation}}
\section{$\!$Another way of derivation of $\bf \frac{S\!O\!(\!2N\!)}{U\!(\!N\!)}\!$
TD Riccati-Hartree-Bogoliubov equation}


~~
As is shown in Section 3,
the matrices $a$ and $b$ and $\chi$
are represented in terms of $q\!=\!(q_{\alpha\beta})$ as\\[-14pt]
\beqa
\left.
\BA{c}
a
\!=\!
(1_{\!N} \!+\! q^\dag q)^{-\frac{1}{2}} 
v ,~
b
\!=\!
q
(1_{\!N} \!+\! q^\dag q)^{-\frac{1}{2}}
v ,~ v \in \! U(N) ,\\
\\[-4pt]
\chi
\!\equiv\!
(1_{\!N} \!+\!  q q^\dag)^{-1} ,~
\bar{\chi}
\!=\!
(1_{\!N} \!+\! q^\dag q)^{-1}  ,
\EA
\right\}
\label{RepmatAandB}
\eeqa
Then the $SO(2N)$ matrix $g$ and the coset variable $q$ are represented as
\beqa
g
=\!
\left[ \!\!
\BA{cc} 
a &~ \bar{b} \\
\\[-2pt]
b &~ \bar{a} \\ 
\EA \!\!
\right]
=\!
\left[ \!\!
\BA{cc} 
a_{0}(\!=\! \bar{\chi}^{\frac{1}{2}})&\bar{b}_{0}(\!=\! \bar{q}\chi^{\frac{1}{2}}) \\
\\[-6pt]
b_{0}(\!=\! q \bar{\chi}^{\frac{1}{2}})&\bar{a}_{0}(\!=\! \chi^{\frac{1}{2}}) \\ 
\EA \!\!
\right] \!\!
\left[ \!\!
\BA{cc} 
v &~ 0 \\
\\[-2pt]
0 &~ \bar{v} \\ 
\EA \!\!
\right] , ~
q
=\!
b a^{-1}
\!=\!
b_{0} a_{0}^{-1} .
\label{gandq}
\eeqa

Along the same strategy as the strategy developed by Chaturvedi et al.
\cite{Chaturvedi.07}
and by Fujii and Oike
\cite{FujiiOike.09},
we here give another way of derivation of $\!\frac{SO(2N)}{U(N)}\!$
time dependent Riccati-Hartree-Bogoliubov (TDRHB) equation.
Let us consider a TD Hamiltonian matrix ${\cal H}(t)$,
$
{\cal H}(t)
\!\!=\!\!
-\bar{\cal F}_F(t)
\!-\!
\bar{\cal F}_D(t)
\!\!=\!\!
\left[ \!\!\!
\BA{cc}
 -\bar{F}(t) &\!\! -\bar{D}(t) \\
\\[-14pt]
D(t) &\!\! F(t)
\EA \!\!\!
\right]
$.
The unitary evolution operator $g(t)$ is an element of
$G \!=\! SO(2N)$ obeying the equation
\beq
i \hbar \dot{g}(t)
\!=\!
{\cal H}(t) g(t), ~
g(t_{0})
\!=\!
\mathbb{I} .
\label{evolequ}
\eeq
Then we have
\beq
i \hbar \!
\left( \!
\left[ \!\!
\BA{cc} 
\dot{a}_{0}& \dot{\bar{b}}_{0} \\
\\[-6pt]
\dot{b}_{0} & \dot{\bar{a}}_{0} \\ 
\EA \!\!
\right] \!\!
\left[ \!\!
\BA{cc} 
v &~ 0 \\
\\[-2pt]
0 &~ \bar{v} \\ 
\EA \!\!
\right]
\!+\!
\left[ \!\!
\BA{cc} 
a_{0}& \bar{b}_{0} \\
\\[-2pt]
b_{0} & \bar{a}_{0} \\ 
\EA \!\!
\right] \!\!
\left[ \!\!
\BA{cc} 
\dot{v} &~ 0 \\
\\[-2pt]
0 &~ \dot{\bar{v}} \\ 
\EA \!\!
\right] \!
\right)
\!\!=\!\!
\left[ \!\!
\BA{cc}
-\bar{F}(t) & -\bar{D}(t) \\
\\[-2pt]
D(t) & F(t)
\EA \!\!
\right] \!\!
\left[ \!\!
\BA{cc} 
a_{0}& \bar{b}_{0} \\
\\[-2pt]
b_{0} & \bar{a}_{0} \\ 
\EA \!\!
\right] \!\!
\left[ \!\!
\BA{cc} 
v &~ 0 \\
\\[-2pt]
0 &~ \bar{v} \\ 
\EA \!\!
\right] ,
\label{evolequ1}
\eeq
which is rewritten as
\beq
i \hbar \!
\left[ \!\!
\BA{cc} 
\dot{a}_{0}& \dot{\bar{b}}_{0} \\
\\[-6pt]
\dot{b}_{0} & \dot{\bar{a}}_{0} \\ 
\EA \!\!
\right] 
\!\!=\!\!
\left[ \!\!
\BA{cc}
-\bar{F}(t) & -\bar{D}(t) \\
\\[-2pt]
D(t) & F(t)
\EA \!\!
\right] \!\!
\left[ \!\!
\BA{cc} 
a_{0}& \bar{b}_{0} \\
\\[-2pt]
b_{0} & \bar{a}_{0} \\ 
\EA \!\!
\right] \! 
\!-\!
i \hbar \!
\left[ \!\!
\BA{cc} 
a_{0}& \bar{b}_{0} \\
\\[-2pt]
b_{0} & \bar{a}_{0} \\ 
\EA \!\!
\right] \!\!
\left[ \!\!
\BA{cc} 
\dot{v} v^{-1}&~ 0 \\
\\[-2pt]
0 &~ \dot{\bar{v}} \bar{v}^{-1}\\ 
\EA \!\!
\right] .
\label{evolequ2}
\eeq
Equation
(\ref{evolequ2})
yields
\beqa
\left.
\BA{c}
i \hbar 
\dot{a}_{0}
\!=\!
-\bar{F}(t) a_{0} \!-\! \bar{D}(t) b_{0}
-
i \hbar
a_{0} \dot{v} v^{-1} , ~
i \hbar 
\dot{b}_{0}
\!=\!
F(t) b_{0} \!+\! D(t) a_{0}
-
i \hbar
b_{0}  \dot{v} v^{-1}  , \\
\\
i \hbar 
\dot{\bar{a}}_{0}
\!=\!
F(t) \bar{a}_{0} \!+\! D(t) \bar{b}_{0}
-
i \hbar 
\bar{a}_{0} \dot{\bar{v}} \bar{v}^{-1} , ~
i \hbar 
\dot{\bar{b}}_{0}
\!=\!
-\bar{F}(t) \bar{b}_{0} \!-\! \bar{D}(t) \bar{a}_{0}
-
i \hbar 
\bar{b}_{0} \dot{\bar{v}} \bar{v}^{-1} .
\EA
\right\} 
\label{evolequ2}
\eeqa
Using the second equation of
(\ref{gandq})
and
(\ref{evolequ2}),
\beqa
\left.
\BA{l}
i \hbar 
\dot{q} 
\!=\!
i \hbar 
\dot{b}_{0} a^{-1}_{0}
\!-\!
i \hbar
b_{0} a^{-1}_{0} \dot{a}_{0} a^{-1}_{0} \\
\\
\!=\!
D(t) \!+\! F(t) q
\!-\!
i \hbar
b_{0}  \dot{v} v^{-1}  a^{-1}_{0}
\!+\!
q \!
\left\{ \!
\bar{F}(t) a_{0} \!+\! \bar{D}(t) b_{0}
\!+\!
i \hbar
a_{0} \dot{v} v^{-1} \!
\right\} \!
a^{-1}_{0}  \\
\\
\!=\!
D(t) \!+\! F(t) q \!+\! q \bar{F}(t)
\!+\! q \bar{D}(t) q .
\EA
\right\}
\label{evolequ3}
\eeqa
This is just the TDRHB equation obtained in
(\ref{RiccatiEq}).
Note that the relation
\beq
\dot{b}_{0} a^{-1}_{0}
\!-\!
b_{0} a^{-1}_{0} \dot{a}_{0} a^{-1}_{0}
\!=\!
\left( \!
\dot{q} \bar{\chi}^{\frac{1}{2}}
\!+\!
{\displaystyle \frac{1}{2}}
q \bar{\chi}^{-\frac{1}{4}} \dot{\bar{\chi}} \bar{\chi}^{-\frac{1}{4}} \!
\right) \!
\bar{\chi}^{-\frac{1}{2}}
\!-\!
q \!
\left( \!
{\displaystyle \frac{1}{2}}
\bar{\chi}^{-\frac{1}{4}} \dot{\bar{\chi}} \bar{\chi}^{-\frac{1}{4}} \!
\right) \!
\bar{\chi}^{-\frac{1}{2}}
\!=\!
\dot{q} ~\!.
\eeq
This is consistent with the second equation of
(\ref{gandq}).


\def\thesection{\Alph{section}}
\setcounter{equation}{0}
\renewcommand{\theequation}{\Alph{section}.\arabic{equation}}
\section{Three- and Four-level models}


~~
In this Appendix, 
we will treat a three-level model $(N \!=\! 3)$.
First we denote the quantities
$q_{12},~q_{13},~q_{23},~
|q^{1}|^{2}
\!\!+\!\!
|q^{2}|^{2}
\!\!+\!\! 
|q^{3}|^{2}$
simply as
$q^{1}(t),~q^{2}(t),~q^{3}(t),~|q|^{2}$, 
respectively.
We also denote
$(\varepsilon_1 \!\!-\!\! \lambda) \!\!+\!\! (\varepsilon_2 \!\!-\!\! \lambda),
~(\varepsilon_1 \!\!-\!\! \lambda) \!\!+\!\! (\varepsilon_3 \!\!-\!\! \lambda),
~(\varepsilon_2 \!\!-\!\! \lambda) \!\!+\!\! (\varepsilon_3 \!\!-\!\! \lambda)$
as
$2 \varepsilon^{1},2 \varepsilon^{2},2 \varepsilon^{3}$
and further
$
D_{12},D_{13}$
and
$D_{23}$
as
${\cal D}^{1}(t) 
\!\!=\!\! 
- s_1 \delta_{1 \bar2 }\Delta(t) ,
\Delta(t)
\!\equiv\!\!
{\displaystyle \frac{1}{2}} g
\left(
s_1 K_{1 \bar1 }(t)
\!+\!
s_2 K_{2 \bar2 }(t)
\!+\!
s_3 K_{3 \bar3 }(t)
\right) , 
$
$\!{\cal D}^{2}(t) 
\!\!=\!\! 
- s_1 \delta_{1 \bar3 }\Delta(t)
$
and
${\cal D}^{3}(t) 
\!\!=\!\! 
- s_2 \delta_{2 \bar3 }\Delta(t)
$, respectively.

The term
$s_1 K_{1 \bar1 }(t)$ is calculated as
\ba
\BA{l}
\\[-20pt]
s_1 K_{1 \bar1 }(t)
\!=\!
s_1 \!\! 
\left[ \!
\left[ \!\!
\BA{ccc}
0 & q^{1} & q^{2} \\
\\[-10pt]
-q^{1} & 0 & q^{3} \\
\\[-10pt]
-q^{2} & -q^{3} & 0
\EA \!\!
\right] \!\!
\left( \!
\left[ \!\!
\BA{ccc}
1 & 0 & 0 \\
\\[-10pt]
0 & 1 & 0 \\
0 & 0 & 1
\EA \!\!
\right]
\!\!-\!\!
\left[ \!\!
\BA{ccc}
0 & \bar{q}^{1} & \bar{q}^{2} \\
\\[-10pt]
-\bar{q}^{1} & 0 & \bar{q}^{3} \\
\\[-10pt]
-\bar{q}^{2} & -\bar{q}^{3} & 0
\EA \!\!
\right] \!\!
\left[ \!\!
\BA{ccc}
0 & q^{1} & q^{2} \\
\\[-10pt]
-q^{1} & 0 & q^{3} \\
\\[-10pt]
-q^{2} & -q^{3} & 0
\EA \!\!
\right] \!
\right)^{\!\!-1}
\right]_{\!\! 1 \bar1} \\
\\[-4pt]
\!=\!
s_1 \!\! 
\left[ \!
\left[ \!\!
\BA{ccc}
0 & q^{1} & q^{2} \\
\\[-10pt]
-q^{1} & 0 & q^{3} \\
\\[-10pt]
-q^{2} & -q^{3} & 0
\EA \!\!
\right] \!\!
\left[ \!\!
\BA{ccc}
1 \!+\! |q^{1}|^{2} \!+\! |q^{2}|^{2} & \bar{q}^{2}q^{3} & -\bar{q}^{1}q^{3} \\
\\[-10pt]
\bar{q}^{3}q^{2} &1  \!+\! |q^{1}|^{2} \!+\! |q^{3}|^{2} & \bar{q}^{1}q^{2} \\
\\[-10pt]
-\bar{q}^{3}q^{1} & \bar{q}^{2}q^{1} & 1  \!+\! |q^{2}|^{2} \!+\! |q^{3}|^{2}
\EA \!\!
\right]^{\!\!-1}
\right]_{\!\! 1 \bar1} \\
\\[-4pt]
\!=\!
s_1 \!\! 
\left[ \!\!
\left[ \!\!
\BA{ccc}
0 & q^{1} & q^{2} \\
\\[-10pt]
-q^{1} & 0 & q^{3} \\
\\[-10pt]
-q^{2} & -q^{3} & 0
\EA \!\!
\right] \!\!
{\displaystyle \frac{1}{1 \!\!+\!\! |q|^{2}}} \!\!\!
\left[ \!\!
\BA{ccc}
1 \!+\! |q^{3}|^{2}&\!\!
-\bar{q}^{2}q^{3}&\!\!
\bar{q}^{1}q^{3} \\
\\[-10pt]
-\bar{q}^{3}q^{2}&\!\!
1 \!+\! |q^{2}|^{2}&\!\!
-\bar{q}^{1}q^{2} \\
\\[-10pt]
\bar{q}^{3}q^{1}&\!\!
-\bar{q}^{2}q^{1}&\!\!
1 \!+\! |q^{1}|^{2}
\EA \!\!
\right] \!\!
\right]_{\!\! 1 \bar1}
\!\!\!=\!
s_1
{\displaystyle \frac{1}{1 \!\!+\!\! |q|^{2}}} \!\!\!
\left[ \!\!
\BA{ccc}
0 & q^{1} & q^{2} \\
\\[-10pt]
-q^{1} & 0 & q^{3} \\
\\[-10pt]
-q^{2} & -q^{3} & 0
\EA \!\!
\right]_{\!\! 1 \bar1} \!\! , 
\EA
\label{K1bar1} 
\ea
where
$|q|^{2}$
is defined as
$|q|^{2}
\equiv\!
|q^{1}|^{2} \!+\! |q^{2}|^{2} \!+\! |q^{3}|^{2}
$.
We also have
\beqa
\BA{l}
s_2 K_{2 \bar2 }(t)
\!=\!
s_2 
{\displaystyle \frac{1}{1 \!\!+\!\! |q|^{2}}} \!\!\!
\left[ \!\!
\BA{ccc}
0 & q^{1} & q^{2} \\
\\[-10pt]
-q^{1} & 0 & q^{3} \\
\\[-10pt]
-q^{2} & -q^{3} & 0
\EA \!\!
\right]_{\!\! 2 \bar2} \!\! , ~~
s_3 K_{3 \bar3 }(t)
\!=\!
s_3 
{\displaystyle \frac{1}{1 \!\!+\!\! |q|^{2}}} \!\!\!
\left[ \!\!
\BA{ccc}
0 & q^{1} & q^{2} \\
\\[-10pt]
-q^{1} & 0 & q^{3} \\
\\[-10pt]
-q^{2} & -q^{3} & 0
\EA \!\!
\right]_{\!\!3 \bar3} \!\! ,
\EA
\label{K2bar2K3bar3}
\eeqa

\vspace{0.1cm}

Using the above quantities,
then we get the following coupled Ricatti equations:
\ba
\BA{ll}
i\hbar \!
\left[ \!\!
\BA{c}
 \dot q^{1}(t) \\
\\[-2pt]
 \dot q^{2}(t) \\
\\[-2pt]
 \dot q^{3}(t) \\
\EA \!\!
\right]
\!=\!\!
\left[ \!\!
\BA{cc}
{\cal D}^{1}(t) \!+\! 2 \varepsilon^{1} \! q^{1}(t)
\!+\!
q^{1}(t) {\cal D}^{1\dag}q^{1}(t)
\!+\!
q^{1}(t) {\cal D}^{2\dag}q^{2}(t)
\!+\!
q^{1}(t) {\cal D}^{3\dag}q^{3}(t)
 \\
\\[-2pt]
{\cal D}^{2}(t) \!+\! 2 \varepsilon^{2} \! q^{2}(t)
\!+\!
q^{1}(t) {\cal D}^{1\dag}q^{2}(t)
\!+\!
q^{2}(t) {\cal D}^{2\dag}q^{2}(t)
\!+\!
q^{2}(t) {\cal D}^{3\dag}q^{3}(t)
 \\
\\[-2pt]
{\cal D}^{3}(t) \!+\! 2 \varepsilon^{3} \! q^{3}(t)
\!+\!
q^{1}(t) {\cal D}^{1\dag}q^{3}(t)
\!+\!
q^{2}(t) {\cal D}^{2\dag}q^{3}(t)
\!+\!
q^{3}(t) {\cal D}^{3\dag}q^{3}(t) 
\EA \!\!
\right] \! .
\EA
\label{RiccatiEqN3}
\ea

\vspace{0.5cm}

Further we treat a four-level model $(N \!\!=\!\! 4)$.
We denote the quantities
$q_{12},q_{13},q_{14},q_{23},q_{24},q_{34}$,
$
|q^{1}|^{2}
\!+\!
|q^{2}|^{2}
\!+\! 
|q^{3}|^{3}
\!+\!
|q^{4}|^{2}
\!+\!
|q^{5}|^{2}
\!+\! 
|q^{6}|^{2}
$
simply as
$q^{1}(t),q^{2}(t),q^{3}(t),q^{4}(t),q^{5}(t),q^{6}(t),|q|^{2}$.\\
Denote
$(\varepsilon_1 \!-\! \lambda) \!+\! (\varepsilon_2 \!-\! \lambda),
(\varepsilon_1 \!-\! \lambda) \!+\! (\varepsilon_3 \!-\! \lambda),
(\varepsilon_1 \!-\! \lambda) \!+\! (\varepsilon_4 \!-\! \lambda)$,
$(\varepsilon_2 \!-\! \lambda) \!+\! (\varepsilon_3 \!-\! \lambda),
(\varepsilon_2 \!-\! \lambda) \!+\! (\varepsilon_4 \!-\! \lambda),
(\varepsilon_3 \!-\! \lambda)\!+\! (\varepsilon_4 \!-\! \lambda)$
as
$2 \varepsilon^{1},2 \varepsilon^{2},2 \varepsilon^{3},
2 \varepsilon^{4},2 \varepsilon^{5},2 \varepsilon^{6}$
and further
$
D_{12},D_{13},D_{14},D_{23},D_{24}$
and
$D_{34}$
as

$~\!{\cal D}^{1}(t)
\!\!=\!\! 
- s_1 \delta_{1 \bar2 }\Delta(t) ,
$, 
$\!{\cal D}^{2}(t)
\!\!=\!\! 
- s_1 \delta_{1 \bar3 }\Delta(t)
$,
$~\!{\cal D}^{3}(t)
\!=\! 
- s_1 \delta_{1 \bar4 }\Delta(t)
$,
${\cal D}^{4}(t)
\!=\! 
- s_2 \delta_{2 \bar3 }\Delta(t)
$,

${\cal D}^{5}(t)
\!=\! 
- s_2 \delta_{2 \bar4 }\Delta(t)
$,
$\!{\cal D}^{6}(t)
\!=\! 
- s_3 \delta_{3 \bar4 }\Delta(t)
$
and

$
\Delta(t)
\!\!\equiv\!\!
{\displaystyle \frac{1}{2}} \! g \!
\left( \!
s_1 K_{1 \bar1 }(t)
\!\!+\!\!
s_2 K_{2 \bar2 }(t)
\!\!+\!\!
s_3 K_{3 \bar3 }(t)
\!\!+\!\!
s_4 K_{4 \bar4 }(t) \!
\right) 
$.

\vspace{0.3cm}

Below we use the
$\mbox{det}$
which is defined as\\[-12pt]
\beqa
\left.
\!\!\!\!
\BA{l}
\mbox{det}
\equiv
\left\{
1 + |q|^{2}
+
|q^{1}|^{2}|q^{6}|^{2} + |q^{2}|^{2}|q^{5}|^{2} + |q^{3}|^{2}|q^{4}|^{2} 
\right. \\
\\[-6pt]
\left.
- \left( \bar{q}^{2}q^{3}q^{4}\bar{q}^{5} + \mbox{c.c} \right)
+ \left( \bar{q}^{1}q^{3}q^{4}\bar{q}^{6} + \mbox{c.c} \right)
+ \left(  \bar{q}^{1}q^{2}q^{5}\bar{q}^{6} + \mbox{c.c} \right) 
\right\}^{2} .
\EA 
\right\}
\eeqa

The term
$s_1 K_{1 \bar1 }(t)$
is computed as\\[-16pt]
\beqa
\!\!\!\!\!\!\!
\BA{l}  
s_1 K_{1 \bar1 }(t)
\!=\!
s_1 \!\! 
\left[ \!
\left[ \!\!
\BA{cccc}
0 &\!\!\! q^{1} &\!\!\! q^{2} &\!\!\! q^{3} \\
\\[-10pt]
-q^{1} &\!\!\! 0 &\!\!\! q^{4} &\!\!\! q^{5} \\
\\[-10pt]
-q^{2} &\!\!\! -q^{4} &\!\!\! 0 &\!\!\! q^{6} \\
\\[-10pt]
-q^{3} &\!\!\! -q^{5} &\!\!\! -q^{6} &\!\!\! 0
\EA \!\!
\right] \!\!
\left( \!
\left[ \!\!
\BA{cccc}
1 & 0 & 0 & 0 \\
\\[-10pt]
0 & 1 & 0 & 0 \\
\\[-10pt]
0 & 0 & 1 & 0 \\
\\[-10pt]
0 & 0 & 0 & 1
\EA \!\!
\right]
\!\!-\!\!
\left[ \!\!
\BA{cccc}
0 &\!\!\! \bar{q}^{1} &\!\!\! \bar{q}^{2} &\!\!\! \bar{q}^{3} \\
\\[-10pt]
-\bar{q}^{1} &\!\!\! 0 &\!\!\! \bar{q}^{4} &\!\!\! \bar{q}^{5} \\
\\[-10pt]
-\bar{q}^{2} &\!\!\! -\bar{q}^{4} &\!\!\! 0 &\!\!\! \bar{q}^{6} \\
\\[-10pt]
-\bar{q}^{3} &\!\!\! -\bar{q}^{5} &\!\!\! -\bar{q}^{6} &\!\!\! 0
\EA \!\!
\right] \!\!
\left[ \!\!
\BA{cccc}
0 &\!\!\! q^{1} &\!\!\! q^{2} &\!\!\! q^{3} \\
\\[-10pt]
-q^{1} &\!\!\! 0 &\!\!\! q^{4} &\!\!\! q^{5} \\
\\[-10pt]
-q^{2} &\!\!\! -q^{4} &\!\!\! 0 &\!\!\! q^{6} \\
\\[-10pt]
-q^{3} &\!\!\! -q^{5} &\!\!\! -q^{6} &\!\!\! 0
\EA \!\!
\right] \!
\right)^{\!\!-1}
\right]_{\!\! 1 \bar1} \\
\\[-2pt]
\!\!\!\!\!\!\!
\!=\!
s_1 \!\! 
\left[ \!
\left[ \!\!
\BA{cccc}
0 &\!\!\! q^{1} &\!\!\! q^{2} &\!\!\! q^{3} \\
\\[-10pt]
-q^{1} &\!\!\! 0 &\!\!\! q^{4} &\!\!\! q^{5} \\
\\[-10pt]
-q^{2} &\!\!\! -q^{4} &\!\!\! 0 &\!\!\! q^{6} \\
\\[-10pt]
-q^{3} &\!\!\! -q^{5} &\!\!\! -q^{6} &\!\!\! 0
\EA \!\!
\right] \!\!\!
\left[ \!\!\!
\BA{cccc}
1 \!\!+\!\! |q^{1}|^{2} \!\!+\!\! |q^{2}|^{2} \!\!+\!\! |q^{3}|^{2}&\!\!\!\!\!
\bar{q}^{2}q^{4} \!+\!  \bar{q}^{3}q^{5}&\!\!\!\!\! -\bar{q}^{1}q^{4} \!+\! \bar{q}^{3}q^{6}
&\!\!\!\!\!
-\bar{q}^{1}q^{5} \!-\! \bar{q}^{2}q^{6}  \\
\\[-10pt]
\bar{q}^{4}q^{2} \!+\! \bar{q}^{5}q^{3}&\!\!\!\!\!
1 \!\!+\!\! |q^{1}|^{2}\!\!+\!\! |q^{4}|^{2} \!\!+\!\! |q^{5}|^{2}
&\!\!\!\!\! \bar{q}^{1}q^{2}  \!+\! \bar{q}^{5}q^{6}
&\!\!\!\!\!
\bar{q}^{1}q^{3} \!-\! \bar{q}^{4}q^{6}  \\
\\[-10pt]
-\bar{q}^{4}q^{1} \!+\! \bar{q}^{6}q^{3}&\!\!\!\!\! \bar{q}^{2}q^{1} \!+\! \bar{q}^{6}q^{5}
&\!\!\!\!\!
1  \!\!+\!\! |q^{2}|^{2} \!\!+\!\! |q^{4}|^{2} \!\!+\!\! |q^{6}|^{2}&\!\!\!\!\!
\bar{q}^{2}q^{3} \!+\!  \bar{q}^{4}q^{5}  \\
\\[-10pt]
-\bar{q}^{5}q^{1} \!-\! \bar{q}^{6}q^{2}&\!\!\!\!\!\bar{q}^{3}q^{1} \!-\! \bar{q}^{6}q^{4}
&\!\!\!\!\!
\bar{q}^{3}q^{2} \!+\!  \bar{q}^{5}q^{4}&\!\!\!\!\!
1  \!\!+\!\! |q^{3}|^{2}\!\!+\!\! |q^{5}|^{2} \!\!+\!\! |q^{6}|^{2}
\EA \!\!\!
\right]^{\!\!-1}
\right]_{\!\! 1 \bar1} \\
\\[-2pt]
\!\!\!\!\!\!\!
\!=\!\!
s_1 \!\!\! 
\left[ 
{\displaystyle \frac{1}{\!\mbox{det}^{\!\frac{1}{2}}}} \!\!\!
\left[ \!\!
\BA{cccc}
0 &\!\!\! q^{1} &\!\!\! q^{2} &\!\!\! q^{3} \\
\\[-10pt]
-q^{1} &\!\!\! 0 &\!\!\! q^{4} &\!\!\! q^{5} \\
\\[-10pt]
-q^{2} &\!\!\! -q^{4} &\!\!\! 0 &\!\!\! q^{6} \\
\\[-10pt]
-q^{3} &\!\!\! -q^{5} &\!\!\! -q^{6} &\!\!\! 0
\EA \!\!
\right] \!\!\!\!
\left[ \!\!\!
\BA{cccc}
1 \!\!+\!\! |q^{4}|^{2} \!\!+\!\! |q^{5}|^{2} \!\!+\!\! |q^{6}|^{2}&\!\!\!\!\!
-\bar{q}^{2}q^{4} \!-\!  \bar{q}^{3}q^{5}&\!\!\!\!\! \bar{q}^{1}q^{4} \!-\! \bar{q}^{3}q^{6}
&\!\!\!\!\!
\bar{q}^{1}q^{5} \!+\! \bar{q}^{2}q^{6}  \\
\\[-10pt]
-\bar{q}^{4}q^{2} \!-\! \bar{q}^{5}q^{3}&\!\!\!\!\!
1 \!\!+\!\! |q^{2}|^{2}\!\!+\!\! |q^{3}|^{2} \!\!+\!\! |q^{6}|^{2}
&\!\!\!\!\! -\bar{q}^{1}q^{2}  \!-\! \bar{q}^{5}q^{6}
&\!\!\!\!\!
-\bar{q}^{1}q^{3} \!+\! \bar{q}^{4}q^{6}  \\
\\[-10pt]
\bar{q}^{4}q^{1} \!-\! \bar{q}^{6}q^{3}&\!\!\!\!\! -\bar{q}^{2}q^{1} \!-\! \bar{q}^{6}q^{5}
&\!\!\!\!\!
1  \!\!+\!\! |q^{1}|^{2} \!\!+\!\! |q^{3}|^{2} \!\!+\!\! |q^{5}|^{2}&\!\!\!\!\!
-\bar{q}^{2}q^{3} \!-\!  \bar{q}^{4}q^{5}  \\
\\[-10pt]
\bar{q}^{5}q^{1} \!+\! \bar{q}^{6}q^{2}&\!\!\!\!\!-\bar{q}^{3}q^{1} \!+\! \bar{q}^{6}q^{4}
&\!\!\!\!\!
-\bar{q}^{3}q^{2} \!-\!  \bar{q}^{5}q^{4}&\!\!\!\!\!
1  \!\!+\!\! |q^{1}|^{2}\!\!+\!\! |q^{2}|^{2} \!\!+\!\! |q^{4}|^{2}
\EA \!\!\!
\right] \!\!
\right]_{\!\! 1 \bar1}  \\
\\[-2pt]
\!\!\!\!\!\!\!
\!=\!
s_1 \!\! 
\left[ 
{\displaystyle \frac{1}{\!\mbox{det}^{\!\frac{1}{2}}}} \!\!\!
\left[ \!\!
\BA{cccc}
0 &\!\!\! q^{1} &\!\!\! q^{2} &\!\!\! q^{3} \\
\\[-10pt]
-q^{1} &\!\!\! 0 &\!\!\! q^{4} &\!\!\! q^{5} \\
\\[-10pt]
-q^{2} &\!\!\! -q^{4} &\!\!\! 0 &\!\!\! q^{6} \\
\\[-10pt]
-q^{3} &\!\!\! -q^{5} &\!\!\! -q^{6} &\!\!\! 0
\EA \!\!
\right] \!
\!+\!
{\displaystyle \frac{1}{\!\mbox{det}^{\!\frac{1}{2}}}} \!\!\!
\left[ \!\!
\BA{cccc}
0 &\!\!\! \bar{q}^{6} &\!\!\! -\bar{q}^{5} &\!\!\! \bar{q}^{4} \\
\\[-10pt]
-\bar{q}^{6} &\!\!\! 0 &\!\!\! \bar{q}^{3} &\!\!\! -\bar{q}^{2} \\
\\[-10pt]
\bar{q}^{5} &\!\!\! -\bar{q}^{3} &\!\!\! 0 &\!\!\! \bar{q}^{1} \\
\\[-10pt]
-\bar{q}^{4} &\!\!\! \bar{q}^{2} &\!\!\! -\bar{q}^{1} &\!\!\! 0
\EA \!\!
\right] \!\!
\left(\! q^{1}q^{6} \!-\! q^{2}q^{5} \!+\! q^{3}q^{4} \! \right) \!
\right]_{\!\! 1 \bar1}, 
\EA
\eeqa
We also have\\[-20pt]
\beqa
\!\!\!\!
\BA{ll}
s_{2(3)} \! K_{2 \bar2 (3 \bar3)} \! (t)
\!\!=\!\!
s_{2(3)} \!\!\!
\left[ 
{\displaystyle \frac{1}{\!\mbox{det}^{\!\frac{1}{2}}}} \!\!\!
\left[ \!\!
\BA{cccc}
0 &\!\!\! q^{1} &\!\!\! q^{2} &\!\!\! q^{3} \\
\\[-10pt]
-q^{1} &\!\!\! 0 &\!\!\! q^{4} &\!\!\! q^{5} \\
\\[-10pt]
-q^{2} &\!\!\! -q^{4} &\!\!\! 0 &\!\!\! q^{6} \\
\\[-10pt]
-q^{3} &\!\!\! -q^{5} &\!\!\! -q^{6} &\!\!\! 0
\EA \!\!
\right] \!
\!\!+\!\!
{\displaystyle \frac{1}{\!\mbox{det}^{\!\frac{1}{2}}}} \!\!\!
\left[ \!\!
\BA{cccc}
0 &\!\!\! \bar{q}^{6} &\!\!\! -\bar{q}^{5} &\!\!\! \bar{q}^{4} \\
\\[-10pt]
-\bar{q}^{6} &\!\!\! 0 &\!\!\! \bar{q}^{3} &\!\!\! -\bar{q}^{2} \\
\\[-10pt]
\bar{q}^{5} &\!\!\! -\bar{q}^{3} &\!\!\! 0 &\!\!\! \bar{q}^{1} \\
\\[-10pt]
-\bar{q}^{4} &\!\!\! \bar{q}^{2} &\!\!\! -\bar{q}^{1} &\!\!\! 0
\EA \!\!
\right] \!\!\!
\left(\! q^{1}q^{6} \!\!-\!\! q^{2}q^{5} \!\!+\!\! q^{3}q^{4} \! \right) \!
\right]_{\!\! 2 \bar2 (3 \bar3)} \!\!\!\! , 
\EA
\eeqa
Then we have the following coupled Ricatti equations:\\[-20pt]
\ba
\BA{ll}
i\hbar \!
\left[ \!\!
\BA{c}
\\[1pt]
\dot q^{1}(t) \\
\\[10pt]
\dot q^{2}(t) \\
\\[10pt]
\dot q^{3}(t) \\
\\[10pt]
\dot q^{4}(t) \\
\\[10pt]
\dot q^{5}(t) \\
\\[10pt]
\dot q^{6}(t) \\
\\[1pt] 
\EA \!\!
\right]
\!=\!\!
\left[ \!\!
\BA{cc}
\\[1pt]
{\cal D}^{1}(t) \!+\! 2 \varepsilon^{1} \! q^{1}(t)
\!+\!
q^{1}(t) {\cal D}^{1 \dag}q^{1}(t)
\!+\!
q^{1}(t) {\cal D}^{2 \dag}q^{2}(t)
\!+\!
q^{1}(t) {\cal D}^{3 \dag}q^{3}(t) \\ 
\!+
q^{1}(t) {\cal D}^{4 \dag}q^{4}(t)
\!+
q^{1}(t) {\cal D}^{5 \dag}q^{5}(t) 
\!+
q^{2}(t) {\cal D}^{6 \dag}q^{5}(t) 
\!+\!
q^{3}(t) {\cal D}^{6 \dag}q^{4}(t) \\
\\[-8pt]
{\cal D}^{2}(t) \!+\! 2 \varepsilon^{2} \! q^{2}(t)
\!+\!
q^{1}(t) {\cal D}^{1 \dag}q^{2}(t)
\!+\!
q^{2}(t) {\cal D}^{2 \dag}q^{2}(t)
\!+\!
q^{2}(t) {\cal D}^{3 \dag}q^{3}(t) \\ 
\!+
q^{2}(t) {\cal D}^{4 \dag}q^{4}(t)
\!+
q^{3}(t) {\cal D}^{5 \dag}q^{4}(t) 
\!+
q^{1}(t) {\cal D}^{5 \dag}q^{6}(t) 
\!+\!
q^{2}(t) {\cal D}^{6 \dag}q^{6}(t) \\
\\[-4pt]
{\cal D}^{3}(t) \!+\! 2 \varepsilon^{3} \! q^{3}(t)
\!+\!
q^{1}(t) {\cal D}^{1 \dag}q^{3}(t)
\!+\!
q^{2}(t) {\cal D}^{2 \dag}q^{3}(t)
\!+\!
q^{3}(t) {\cal D}^{3 \dag}q^{3}(t) \\
\!-
q^{1}(t) {\cal D}^{4 \dag}q^{6}(t)
\!+
q^{2}(t) {\cal D}^{4 \dag}q^{5}(t) 
\!+
q^{3}(t) {\cal D}^{5 \dag}q^{5}(t) 
\!+\!
q^{3}(t) {\cal D}^{6 \dag}q^{6}(t) \\
\\[-4pt]
{\cal D}^{4}(t) \!+\! 2 \varepsilon^{4} \! q^{4}(t)
\!+\!
q^{1}(t) {\cal D}^{1 \dag}q^{4}(t)
\!+\!
q^{2}(t) {\cal D}^{2 \dag}q^{4}(t)
\!-\!
q^{1}(t) {\cal D}^{3 \dag}q^{6}(t) \\
\!+
q^{2}(t) {\cal D}^{3 \dag}q^{5}(t)
\!+
q^{4}(t) {\cal D}^{4 \dag}q^{4}(t) 
\!+
q^{4}(t) {\cal D}^{5 \dag}q^{5}(t) 
\!+\!
q^{4}(t) {\cal D}^{6 \dag}q^{6}(t) \\
\\[-4pt]
{\cal D}^{5}(t) \!+\! 2 \varepsilon^{5} \! q^{5}(t)
\!+\!
q^{1}(t) {\cal D}^{1 \dag}q^{5}(t)
\!+\!
q^{1}(t) {\cal D}^{2 \dag}q^{6}(t)
\!+\!
q^{3}(t) {\cal D}^{2 \dag}q^{4}(t) \\
\!+
q^{3}(t) {\cal D}^{3 \dag}q^{5}(t)
\!+
q^{4}(t) {\cal D}^{4 \dag}q^{5}(t) 
\!+
q^{5}(t) {\cal D}^{5 \dag}q^{5}(t) 
\!+\!
q^{5}(t) {\cal D}^{6 \dag}q^{6}(t) \\
\\[-4pt]
{\cal D}^{6}(t) \!+\! 2 \varepsilon^{6} \! q^{6}(t)
\!+\!
q^{2}(t) {\cal D}^{1 \dag}q^{5}(t)
\!-\!
q^{3}(t) {\cal D}^{1 \dag}q^{4}(t)
\!+\!
q^{2}(t) {\cal D}^{2 \dag}q^{6}(t) \\
\!+
q^{3}(t) {\cal D}^{3 \dag}q^{6}(t)
\!+
q^{3}(t) {\cal D}^{4 \dag}q^{6}(t) 
\!+
q^{5}(t) {\cal D}^{5 \dag}q^{6}(t) 
\!+\!
q^{6}(t) {\cal D}^{6 \dag}q^{6}(t) 
\\[8pt]
\EA \!\!
\right] \! .
\EA
\label{RiccatiEqN4}
\ea


\newpage

\vskip0.8cm
\begin{center}
{\bf Acknowledgements}
\end{center}
S. Nishiyama would like to
express his sincere thanks to
Professor Constan\c{c}a Provid\^{e}ncia for kind and
warm hospitality extended to
him at the Centro de F\'\i sica,
Universidade de Coimbra, Portugal.
This work was supported by FCT (Portugal) under the project
CERN/FP/83505/2008.
The authors thank the Yukawa Institute for Theoretical Physics
at Kyoto University. Discussions during the YITP workshop
YITP-W-09-04 on ``Development of Quantum Field Theory and String Theory''
were useful to complete this work.

\newpage



\begin{thebibliography}{99}
\vspace{-0.1cm}
\bibitem{Zumino.79}
B. Zumino,
Supersymmetry and K\"{a}hler manifolds,
Phys. Lett. {\bf 87B} (1979) 203-206.
\bibitem{NNH.01}
S. Groot Nibbelink, T.S. Nyawelo and J.W. van Holten,
Construction and analysis of anomaly-free
supersymmetric $SO(2N)/U(N) ~\sigma$-models,
Nucl. Phys. {\bf B594} (2001) 441-476.
\bibitem{Bog.59}
N. N. Bogoliubov,
The compensation principle and the self-consistent field method,
Soviet Phys. Uspekhi, {\bf 67} (1959) 236-254.
\bibitem{RS.80}
P. Ring and P. Schuck,
{\it The nuclear many-body problem}, Springer, Berlin, 1980.
\bibitem{BlaizotRipka.86}
J.P. Blaizot and G. Ripka,
{\it Quantum Theory of Finite Systems},
MIT Press, Cambridge, MA, 1986.
\bibitem{Perelomov.86}
A.M. Perelomov,
{\it Generalized Coherent States and Their Applications},
Springer-Verlag, 1986;\\
Chiral models:Geometrical aspects,
Phys. Rep. {\bf 146} (1987) 135-213 and references therein.
\bibitem{FYN.77} 
H. Fukutome, M. Yamamura and S. Nishiyama,
A new fermion many-body theory based on the $SO(2N+1)$ Lie algebra
of the fermion operators,
Prog. Theor. Phys. {\bf 57} (1977) 1554-1571.
\bibitem{YN.76}
M. Yamamura and S. Nishiyama,
An a priori quantized time-dependent Hartree-Bogoliubov theory. 
- A generalization of the Schwinger representation of 
quasi-spin to the fermion pair algebra -,
Prog. Theor. Phys. {\bf 56} (1976) 124-134.
\bibitem{SJCF.08}
Seiya Nishiyama, Jo\~ao da Provid\^{e}ncia,
Constan\c{c}a Provid\^{e}ncia and Fl\' avio Cordeiro,
Extended supersymmetric $\sigma$-model based on the $SO(2N+1)$ Lie algebra
of the fermion operators,
Nucl. Phys. {\bf B802} (2008) 121-145.
\bibitem{Fuk.77}
H. Fukutome,
On the SO(2N+1) regular representation of operators and wave functions of fermion many-body systems,
Prog. Theor. Phys. {\bf 58} (1977) 1692-1708;\\
H. Fukutome and S. Nishiyama, 
Time dependent $SO(2N+1)$ theory for unified description of bose and fermi type collective excitations,
Prog. Theor. Phys. {\bf 72} (1984) 239-251.
\bibitem{Fuk.81}
H. Fukutome,
The group theoretical structure of fermion many-body systems arising from the canonical anticommutation relation. I
{\it- Lie algebras of fermion operators and exact generator coordinate representations
of state vectors -},
Prog. Theor. Phys. {\bf 65} (1981) 809-827.
\bibitem{Nishi.81}
S. Nishiyama,
Path integral on the coset space of the $SO(2N)$ group and the time-dependent
Hartree-Bogoliubov equation,
Prog. Theor. Phys. {\bf 66} (1981) 348-350.
\bibitem{BG.91}
V. Ceausescu and A. Gheorghe,
Classical limit and quantization of Hamiltonian systems,
in
{\it Symmetries and Semiclassical Features of Nuclear Dynamics},
Invited Lectures of the 1986 International Summer School,
Edited by A. A. Raduta,
Lecture Notes in Physics, {\bf 279}, 69-117, 
Springer-Verlag Berlin Heiderberg 1987.\\
S. Berceanu and A. Gheorghe,
On equations of motion on complex Grassmann manifold,
Rev. Roum. Phys. {\bf 36} (1991) 533-554;\\
On equations of motion on compact Hermitian symmetric spaces,
J. Math. Phys. {\bf 33} (1992) 998-1007.\\
S. Berceanu and L. Boutet de Monvel,
Linear dynamical systems, coherent state manifolds, flows, and
matrix Riccati equation,
J. Math. Phys. {\bf 34} (1993) 2353-2371.
\bibitem{Riccati.1758}
Count J.F. Riccati,
Animadversationes in aequationes differentiales secundi gradus,
{\it Actorum Eruditorum quae Lipsiae publicantur.  Supplementa}
 {\bf 8} (1724) 66-73. 
\bibitem{Reid.72}
T.W. Reid,
{\it Riccati Differential Equation},16
Mathematices in Science and Engineering, Vol. {\bf 86}
(Academic, New York, 1972).
\bibitem{Z_Itkin.73}
Zakhar-Itkin,
The matrix Riccati differential equation and the semi-group of 
linear fractional transformations,
Russ. Math. Surv. {\bf 28}:3 (1973) 89-131;
Uspekhi Mat. Nauk, {\bf 28}:3 (1973) 83-120. 
\bibitem{BKMU.84}
M. Bando, T. Kuramoto, T. Maskawa and S. Uehara,
Structure of non-linear realization in supersymmetric theories,
Phys. Lett. {\bf B} 138 (1984) 94-98;\\
Non-linear realization in supersymmetric theories,
Prog. Theor. Phys. {\bf 72} (1984) 313-349;
Non-linear realization in supersymmetric theories. II,
{\it ibid} 1207-1213.\\
M. Bando, T. Kugo and K. Yamawaki,
Nonlinear realization and hidden local symmetries,
Phys. Rep. {\bf 164} (1988) 217-314.
\bibitem{SJCF.11}
Seiya Nishiyama, Jo\~ao da Provid\^{e}ncia,
Constan\c{c}a Provid\^{e}ncia and Fl\' avio Cordeiro,
Anomaly-free supersymmetric $SO(2N+2)/U(N+1) ~\sigma$-model
based on the $SO(2N+1)$ Lie algebra of the fermion operators,
{\it JHEP} {\bf02} (2011) 093-1-093-30.
\bibitem{Inoguchi.01}
Jun-ichi Inoguchi,
{\it Secret of Riccati}, in Japanese, Nihon Hyouron Sha Company, 2010.
\bibitem{Chaturvedi.07}
S. Chaturvedi, E. Ercolessi, G. Marmo, G. Morandi, N. Mukunda and R. Simon,
Ray space 'Riccati' evolution and geometric phases for $N$-level quantum spaces,
Pramana J. Phys. {\bf 69} (2007) 317-328;arXiv:0706.0964 [quanta-ph].
\bibitem{FujiiOike.09}
K. Fujii and H. Oike,
Reduced dynamics from the unitary group to some flag manifolds: Interacting matrix equations,
Int. J. Geom. Methods Mod. Phys. {\bf 6} (2009) 573-581; arXiv:0809.0165 [math-ph].
\end{thebibliography}
\end{document}